\documentclass[journal=jacsat,manuscript=article]{achemso}

\usepackage[version=3]{mhchem} 

\usepackage{graphicx}
\usepackage{bm}

\usepackage[usenames,dvipsnames]{color}
\usepackage{braket}
\usepackage{comment}
\usepackage[english]{babel}
\usepackage{wasysym}
\usepackage[colorlinks,bookmarks=false,citecolor=blue,linkcolor=red,urlcolor=blue]{hyperref}

\usepackage{CJK}

\author{Chong Ye}
\affiliation{Beijing Key Laboratory of Nanophotonics and Ultrafine Optoelectronic Systems, School of Physics, Beijing Institute of Technology, 100081 Beijing, China}
\author{Yifan Sun}
\affiliation{Beijing Key Laboratory of Nanophotonics and Ultrafine Optoelectronic Systems, School of Physics, Beijing Institute of Technology, 100081 Beijing, China}
\author{Yong Li}\email{yongli@hainanu.edu.cn}
\affiliation{Center for Theoretical Physics, Hainan University, Haikou 570228, China}
\author{Xiangdong Zhang}\email{zhangxd@bit.edu.cn}
\affiliation{Beijing Key Laboratory of Nanophotonics and Ultrafine Optoelectronic Systems, School of Physics, Beijing Institute of Technology, 100081 Beijing, China}

\title[An \textsf{achemso} demo]
  {
  Single-shot Non-destructive Quantum Sensing for Gaseous Samples with Hundreds of Chiral Molecules}

\abbreviations{IR,NMR,UV}

\begin{document}


\begin{abstract}
{Chiral discrimination that is efficient to tiny amounts of chiral substances, especially at the single-molecule level, is highly demanded. Here, we propose a single-shot non-destructive quantum sensing method addressing such an issue}. Our scheme consists of two steps. In the first step, the two enantiomers are prepared in different rotational states via microwave enantio-specific state transfer. Then, the chiral discrimination is transferred to quantum hypothesis testing. In the second step, we for the first time introduce a non-destructive quantum-state detection technique assisted with a microwave resonator to chiral discrimination, through which the molecular chirality is determined by the sign of the output signals. Using a typical chiral molecule, 1,2-propanediol, and an experimentally feasible model based on spherical Fabry-P\'{e}rot cavity, we show that the molecular chirality of slowly moving enantiopure gaseous samples with $10^2\sim10^3$ molecules can be highly credibly distinguished in a single-shot detection. By further trapping chiral molecules, it is promising to achieve chiral discrimination at the single-molecule level by using our approach. 
\end{abstract}


\begin{tocentry}
\begin{minipage}{ 4.45cm}
\includegraphics[width=1.5\columnwidth]{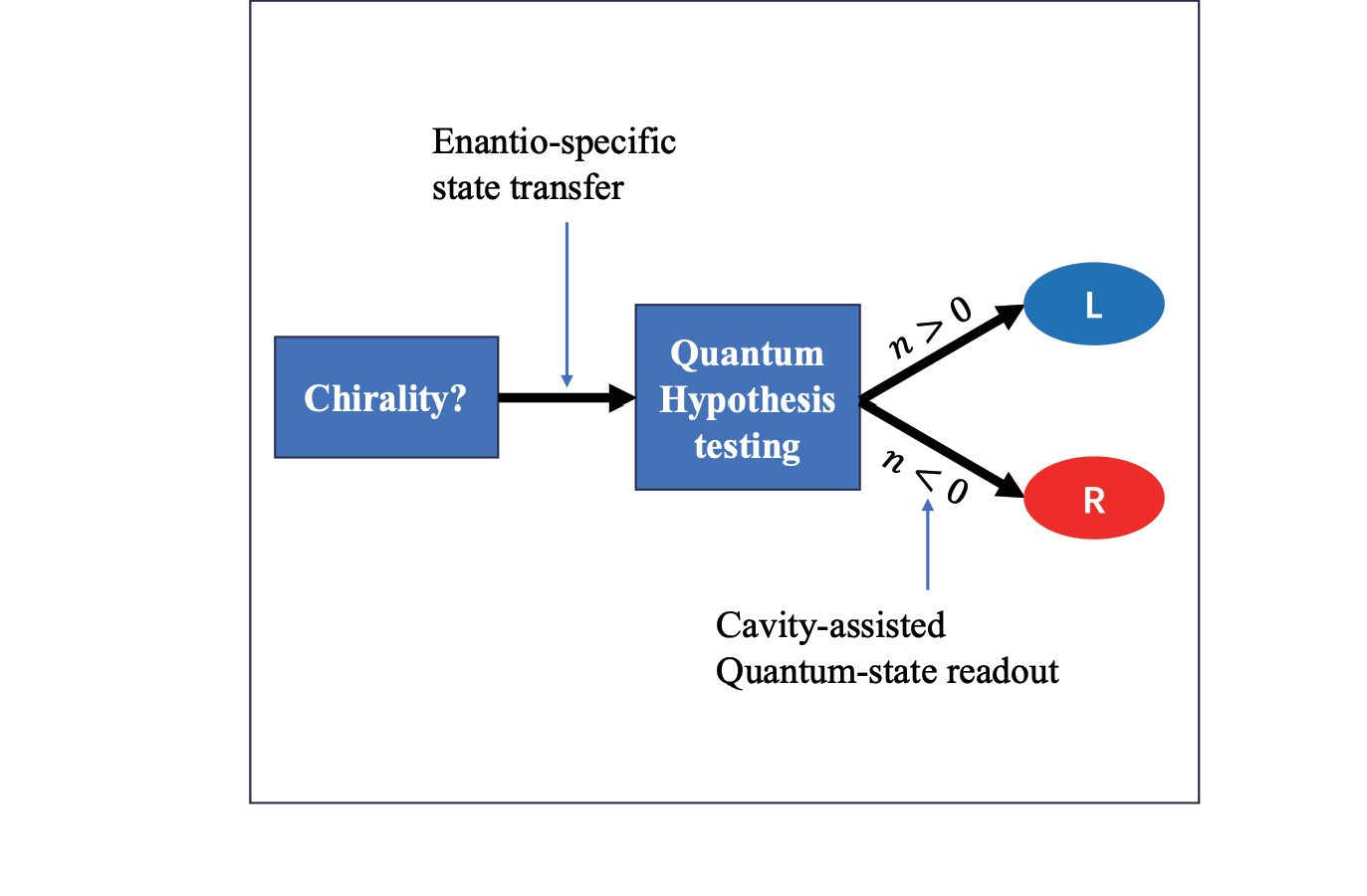}
\end{minipage}
\end{tocentry}

Chiral molecules, also known as enantiomers, are mirror images of each other but are not super-imposable by translations and rotations. Their significance is emphasized by the enantioselectivity in the activity of biological molecules and broad classes of chemical reactions as well as the homochirality in life. Chiral discrimination, distinguishing two enantiomers of opposite chiralities, is a vitally important problem since the first discovery of molecular chirality~\cite{pasteur1915researches}. According to Curie's dissymmetry principle, two enantiomers can be distinguished by interacting with another symmetry-broken (chiral) item, including light, molecule, and media. In traditional chiroptical methods, such as optical rotary dispersion, circular dichroism, and Raman optical activity, circularly polarized light was used for chiral discrimination. These methods relied on relatively weak optical magnetic interactions, such that highly concentrated samples with a large number of chiral molecules are needed. In addition, circularly polarized fields were also proposed to obtain the enantio-specific state transfer~\cite{salam1997control} and the laser-induced asymmetric synthesis~\cite{shao1997control}. 

For biomedical and pharmaceutical applications, ultra-sensitive chiral discrimination methods efficient to tiny amounts of chiral substances are highly demanded. {The achievement of chiral discrimination at the single-molecule level, such as the chiral scanning tunneling microscopy on surface-bond molecules~\cite{tierney2011regular,balema2021enantioselective}, can substantially promote the related areas.} Recently, the chiral discrimination in purely electric dipole effect was achieved in cold gaseous samples by using the microwave three-wave mixing technique~\cite{patterson2013enantiomer,patterson2013sensitive,patterson2014new,shubert2014identifying,shubert2015rotational,lobsiger2015molecular,shubert2016chiral,grabow2013fourier}, {where two microwave fields are applied to generate a third field. Because the transition electric dipoles change signs with two enantiomers, the phase of the third field differs by $\pi$ for the two enantiomers.} {This opens up a new avenue for developing ultra-sensitive chiroptical methods.} An enantiomeric excess resolution of $2\%$ by using $10^6$ repeated measurements in cold gaseous samples of $10^{11}$ effective working molecules was reported~\cite{patterson2013sensitive}. {The enantio-specific state transfer~\cite{eibenberger2017enantiomer,perez2017coherent,lee2022quantitative} is a direct extension of microwave three-wave mixing. The two enantiomers initially in the same-energy inner state can be transferred to different-energy inner states by applying three microwave fields.} Combining it with a state-selective free-induced decay (FID) detection, an enantiomeric excess resolution of $0.05\%$ was realized~\cite{eibenberger2017enantiomer}. Further, some sensitive state-selective detection techniques, including laser-induced fluorescence (LIF), resonance-enhanced multiphoton ionization (REMPI), resonance-enhanced multiphoton dissociation (REMPD), and buffer-gas infrared absorption spectroscopy (BGIRAS) were suggested to enhance the resolution~\cite{eibenberger2017enantiomer}. Very recently, the LIF scheme was experimentally demonstrated~\cite{lee2022quantitative}.

However, the commonly state-selective detection methods have severe drawbacks {for tiny amounts of chiral substances} due to their destructive feature. For LIF, REMPI, and REMPD, there are many competing processes due to the complex level structure of chiral molecules. Then, the desired signals therein are usually not sufficiently generated before the initial molecular states are destroyed. For BGIRAS, it gets the frequency information by detecting signals at different locations, such that a single absorption event does not provide enough information about the spectrum. Moreover, the chiral molecules may be destroyed or disturbed after destructive detection, resulting in the chemically and biologically important chiral molecules ceasing to be effective. 

\begin{figure}[htp]
	\centering
	\includegraphics[width=0.85\columnwidth]{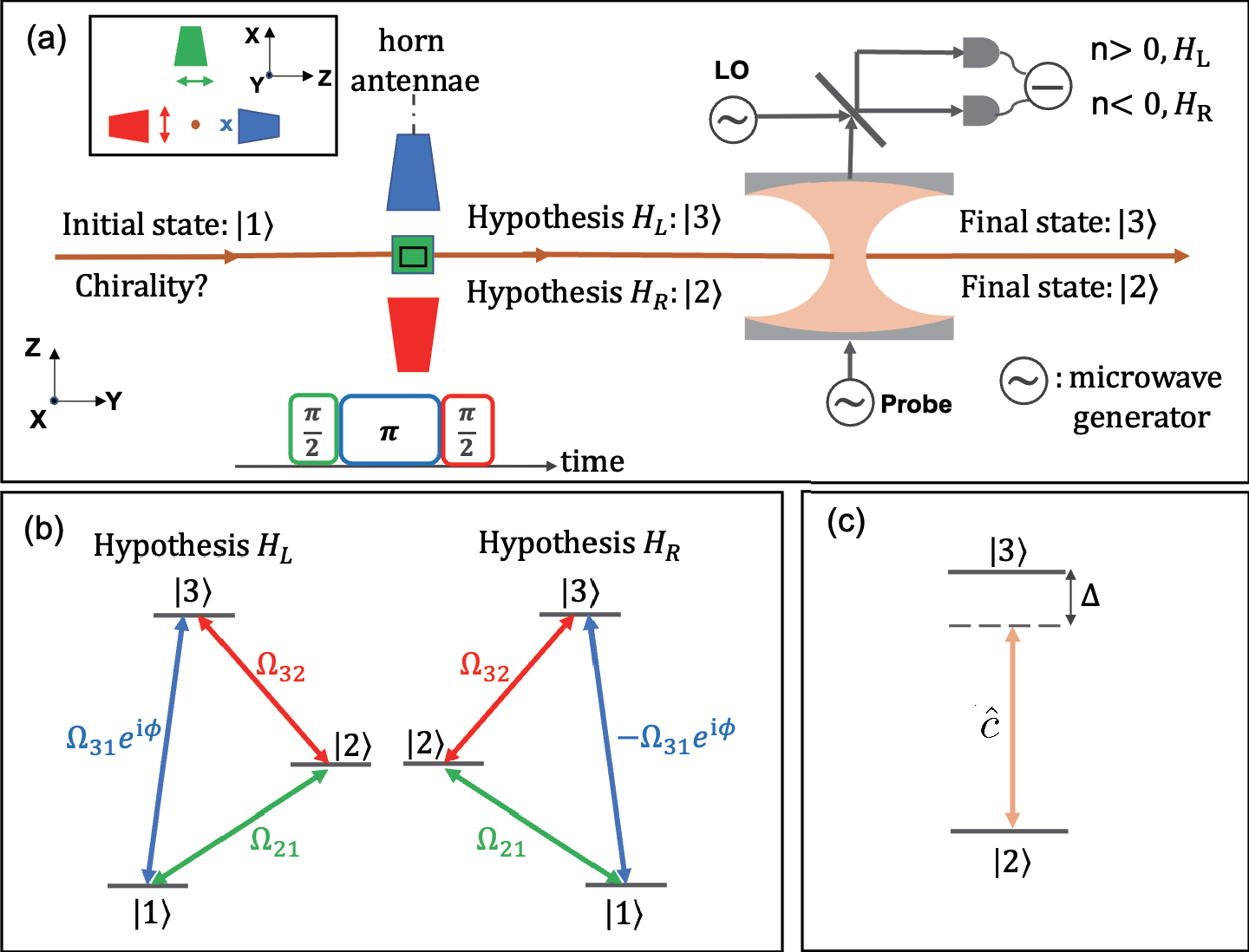}\\
	\caption{(a) Scheme of the chiral quantum sensor. In step I (the left panel), the tricolor synthetic light composed of three microwave pulses with mutually orthogonal polarization directions (the left upper corner) is applied through the horn antennas to generate the enantioselective states. The two enantioselective states serve as two quantum hypotheses, which are determined by monitoring the enantioselective responses of the cavity mode via homodyne detection in step II (the right panel). The microscopic light-molecule interaction models for the two steps are shown in (b) and (c).}\label{Fig1}
\end{figure}

In this Letter, we theoretically propose {an ultra-sensitive single-shot non-destructive quantum sensor for chiral discrimination of gaseous samples with hundreds of chiral molecules.} It consists of two steps as shown in Fig.\,\ref{Fig1}(a). We first prepare chiral molecules of opposite chirality to different rotational states by microwave enantio-specific state transfer, transferring the chiral discrimination to quantum hypothesis testing. Then, we for the first time introduce the dispersive detection technique assisted with a microwave resonator, where a balanced Homodyne measurement is used and the molecule chirality is distinguished by the sign of the output signals. {Our method is ultra-sensitive, without the requirement of repeated measurement, and non-destructive, offering advantages over three-wave mixing technique~\cite{patterson2013enantiomer,patterson2013sensitive,patterson2014new,shubert2014identifying,shubert2015rotational,lobsiger2015molecular,shubert2016chiral,grabow2013fourier}.} To illustrate these advantages, we provide numerical simulations by using a typical chiral molecule, 1,2-propanediol, and an experimentally feasible model based on a spherical Fabry-P\'{e}rot cavity.  The numerical results show that the molecular chirality of slowly moving enantiopure samples with $10^2\sim10^3$ molecules can be highly credibly distinguished in a single-shot detection. It is promising to achieve chiral discrimination at the single-molecule level by further trapping chiral molecules.

\textit{Cavity-assisted chiral quantum sensor.}
As shown in Fig.\,\ref{Fig1}(a), we assume that chiral molecules with unknown chirality are initially populated in the state $|1\rangle$. In step I, we realize the enantio-specific state transfer of chiral molecules by applying three pulses with well-designed polarization and frequencies~\cite{ye2018real} [see the left upper corner of Fig.\,\ref{Fig1}(a)], such that the working model of step I is described by the cyclic three-level model [see Fig.\,\ref{Fig1}(b)]. The working Hamiltonian of step I in the interaction picture is
\begin{align}\label{TCM}
    \hat{H}_{Q}=\sum^{3}_{j>i=1}\Omega^{Q}_{ij}(t)|i\rangle\langle j|+h.c.,
\end{align}
where subscript $Q=(L,R)$ denotes the molecular chirality and the enantioselective coupling strengths $\Omega^{Q}_{ij}$ are
\begin{align}
    \Omega^{Q}_{21}=\Omega_{21},~~ \Omega^{Q}_{32}=\Omega_{32}, ~~\Omega^{L}_{31}=-\Omega^{R}_{31}=\Omega_{31}e^{\mathrm{i}\phi}.
\end{align}
For simplicity and without loss of generality, $\Omega_{ji}$ are assumed to be positive. The cyclic three-level model offers the possibility of generating enantioselectivity in purely electric dipole effect, and thus intensely discussed in chiral discrimination and enantiopurification~\cite{kral2001cyclic,kral2003two,li2007generalized,vitanov2019highly,ye2021entanglement,khokhlova2022enantiosensitive,cai2022enantiodetection,leibscher2022full}. {In our treatment, we have also neglected the predicted tiny energy differences between the two enantiomers due to parity-violating interactions~\cite{quack2012molecular}.}


We apply three non-overlapping resonant pulses [see the left panel of Fig.\,\ref{Fig1}(a)], yielding the molecule process in the following manner~\cite{lee2022quantitative} $|2\rangle\xleftarrow{\pi/2}|1\rangle\xrightarrow{\pi}|3\rangle\xrightarrow{\pi/2}|2\rangle$. The overall phase is adjusted to $\phi=-\pi/2$ by tuning the phases of the applying fields. The final states of the molecules are enantioselective [see the middle panel of Fig.\,\ref{Fig1}(a)]. Then, the distinguishing of molecular chirality is transferred to testing the two most distinguishable quantum hypotheses~\cite{spedalieri2014asymmetric} 
\begin{align}\label{EH}
   &\mathrm{Hypothesis}~{H}_L: |\Psi\rangle_{{H}_L}=|3\rangle,\nonumber\\
   &\mathrm{Hypothesis}~{H}_R: |\Psi\rangle_{{H}_R}=|2\rangle.
\end{align}
For details about the working states of the cyclic three-level model, the origin of its enantioselectivity in pure electric dipole-interaction physics, and the evolution of the chiral molecule, see Sec.\,1.1 in Supporting Information.  
 
In step II, we determine the two hypotheses in Eq.\,(\ref{EH}) with the help of a driven microwave cavity as shown in the right panel of Fig.\,\ref{Fig1}(a). 
{We constantly and resonantly drive the cavity mode $\hat{c}$ with a classical $\omega_0$-field generated by the microwave generator labeled `Probe' in Fig.\,\ref{Fig1}(a). The geometry of the cavity is well designed, such that the cavity mode $\hat{c}$ near-resonantly couples with the rotational transition $|2\rangle\leftrightarrow|3\rangle$ as shown in Fig.\,\ref{Fig1}(c). The hybrid system with $N_m$ molecules can be described by the driven Tavis–Cummings model ($\hbar=1$)
\begin{align}\label{HMD}
    \hat{\mathcal{H}}=\sum^{N_{m}}_{l=1}\Delta_{m}\hat{\sigma}^{\dagger}_{l}\hat{\sigma}_{l}+\sum^{N_{m}}_{l=1} \bar{g}(\hat{c}\hat{\sigma}^{\dagger}_{l}+
    \hat{c}^{\dagger}\hat{\sigma}_{l})+\mathrm{i}\eta(\hat{c}^{\dagger}-\hat{c}).
\end{align}
The operators for the $l$-th molecule are defined as $\hat{\sigma}_{l}=|2\rangle_{ll}\langle 3|$ and $\hat{\sigma}^{z}_{l}=2\hat{\sigma}^{\dagger}_{l}\hat{\sigma}_{l}-1$. The detunings are $\Delta_c=\omega_c-\omega_0=0$ and $\Delta_m=\omega_{32}-\omega_0$. Then, the cavity-molecule detuning is $\Delta=\Delta_m-\Delta_c=\Delta_m$. The cavity pumping rate is $\eta$. We consider the case that the size of the sample is much smaller than the width of the spatial profile of the cavity field, such that the coupling strength is assumed identical for different molecules in the sample and $\bar{g}$ is used in Eq.\,(\ref{HMD}) (for more details, see Sec.\,2 of Supporting Information).

{We assume that the free space decay, the collective decay, and the dipole-dipole interaction between molecules are negligible}, yielding the equations of motion for $\sigma\equiv \sum^{N_m}_{l=1}\langle \hat{\sigma}_l\rangle/N_m$ and ${\sigma}^z\equiv\sum^{N_m}_{l=1}\langle\hat{\sigma}^z_l\rangle/N_m$ according to first-order mean-field theory in the cumulant expansion~\cite{kubo1962generalized} (for more details, see Sec.\,2.1 of Supporting Information)
\begin{align}\label{SS1}
&\dot{ {c}}=-\kappa {c}-\mathrm{i}\bar{g}N_m {\sigma}+\eta,\nonumber\\
&\dot{ {\sigma}}=-\mathrm{i}\Delta_m {\sigma}+\mathrm{i}\bar{g}{c} \sigma^z,\nonumber\\
&\dot{ {\sigma}}^z=2\mathrm{i}\bar{g}( {c}^{ \ast} {\sigma}- {c} {\sigma}^{ \ast}).
\end{align}
The cavity decay rate is $\kappa$. 
When the molecular sample moves sufficiently slowly, the state-selective responses of the cavity mode are  
$c(t)\simeq {\eta}\exp{[i\varphi_Q(t)]}/{\kappa}$ with hypothesis-selective phase shift
\begin{align}\label{SB1}
\varphi_{L}(t)=-\varphi_{R}(t)=-\frac{\bar{g}^2(t)N_m}{\kappa\Delta_m}.
\end{align} 
Here, we have chosen the initial state of cavity mode to be $c(0)=\eta/\kappa$. {We note that $\bar{g}$ becomes time-dependent due to the movement of the sample. For the deduction of Eq.\,(\ref{SB1}) in the dispersive limit, see Sec.\,3.3 of Supporting Information. 
} 

To monitor the state-selective responses of cavity mode, homodyne detection is established by applying a strong local field as sketched in the upper right panel of Fig.\,\ref{Fig1}(a). The local field is generated by the microwave generator labeled `LO', and is mixed with the output of the cavity by a $50:50$ beam splitter. Two microwave photon counters or a digital receiver are used to monitor the two output ports of the beam splitter, whose difference provides our final signals. 

\begin{figure}[hpt]
	\centering
	\includegraphics[width=0.85\columnwidth]{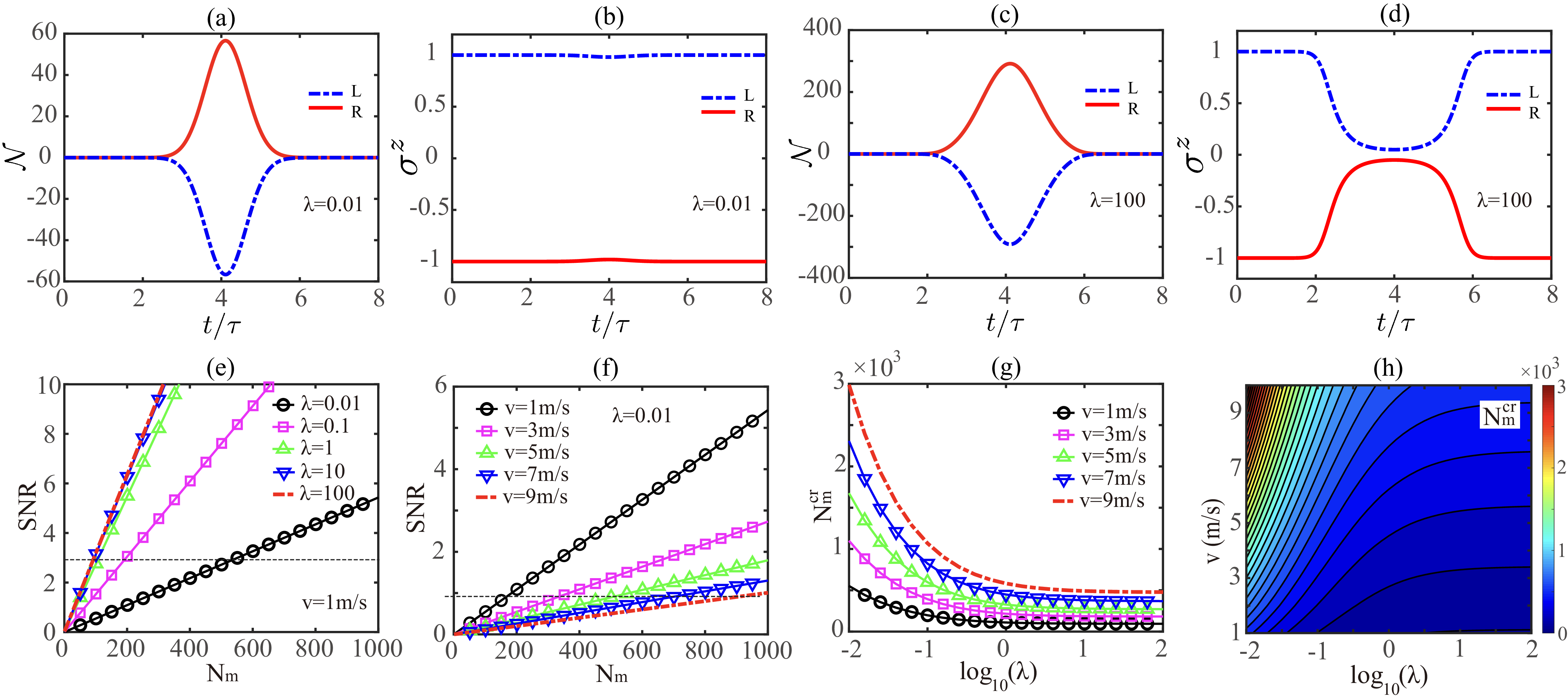}\\
	\caption{Numerical simulations for 1,2-propanediol. (a-d) give the time evolutions of interest $\mathcal{N}$ (in the unit of $\sqrt{\mathrm{Hz}}$) and $\sigma^z$. (a) and (b) are in the dispersive case with $\lambda\equiv N_0/N_{\mathrm{cr}}=0.01$. (c) and (d) are in the non-dispersive case with $\lambda=100$. Here, $N_0=\eta^2/\kappa^2$ is the photon number corresponding to the initial state of the cavity mode with $c(0)=\eta/\kappa$ and $N_{\mathrm{cr}}$ is the critical photon number of the dispersive region. The particle number and the forward velocity of the samples are $N_m=1000$ and $\mathrm{v}=1$\,$\mathrm{m/s}$. The time unit is $\tau\equiv w_0/\mathrm{v}$. Here, it is about $11.59$\,$\mathrm{ms}$. (e) and (f) give the signal-to-noise rate ($\mathrm{SNR}$) of single-shot measurement as functions of $N_m$ for different $\lambda$ and $\mathrm{v}$. The dashed horizontal lines indicate $\mathrm{SNR}=3$ with the corresponding error probability of $P_{\mathrm{err}}\simeq0.001$, whose intersections with the functions of $N_m$ give the critical values of $N_m$ (i.e., $N^{\mathrm{cr}}_m$) for highly credible distinguishing of molecular chirality. (g) shows $N^{\mathrm{cr}}_m$ as functions of $\lambda$ for different $\mathrm{v}$. (h) shows $N^{\mathrm{cr}}_m$ on the $\lambda-\mathrm{v}$ plane for slowly moving enantiopure samples. }\label{Fig2}
\end{figure}

Then, a single-shot detection, collecting signals from $t_0$ to $t_f$, yields a Gaussian random variable $n_Q$ with the standard derivation $\delta=\sqrt{N_{\mathrm{lo}}(t_f-t_0)}$~\cite{horak2003possibility} and the enantioselective mean value
\begin{align}\label{EnQ}
\bar{n}_Q
\simeq\int^{t_f}_{t_0}\sqrt{2\kappa }\Re[{c}^{\ast}_{\mathrm{lo}}c(t^{\prime})]dt^{\prime}\equiv\int^{t_f}_{t_0}\mathcal{N}(t^{\prime})|c_{\mathrm{lo}}| dt^{\prime}.
\end{align}
Here, we have defined $\mathcal{N}\equiv \sqrt{2\kappa }\Re({c}^{\ast}_{\mathrm{lo}}c)/|{c}_{\mathrm{lo}}|$.
When the phase of the local field is $\varphi_{\mathrm{lo}}=\pm\pi/2$, the mean values $\bar{n}_Q$ change signs with hypotheses, i.e., the hypotheses can be decided by the sign of the counting difference between the two detectors depicted in the upper right corner of Fig.\,\ref{Fig1}(a). For more details about the detection scheme, see Sec.\,3 of Supporting Information.


\textit{Numerical simulations for 1,2-propanediol.}
We take a typical chiral molecule, 1,2-propanediol, as an example to illustrate our scheme. The working states are chosen as~\cite{ye2018real}  $|1\rangle=|0_{0,0,0}\rangle$, 
$|2\rangle=|1_{0,1,0}\rangle$, and $|3\rangle=(|1_{1,0,1}\rangle+|1_{1,0,-1}\rangle)/\sqrt{2}$, where the rotational eigenstates are given in $|J_{K_a,K_c,M}\rangle$ notation. The angular frequencies of the three 
microwave fields are~\cite{ye2018real} $\omega_{21}=2\pi\times 6.43106$\,GHz, $\omega_{31}=2\pi\times 12.21215$\,GHz, and $\omega_{32}=2\pi\times 5.781096$\,GHz. We choose the $\mathrm{TEM}_{000}$ mode of the well-designed spherical Fabry-P\'{e}rot cavity as the working cavity mode. The cavity decay rate is about $\kappa=2\pi\times121.7$\,$\mathrm{Hz}$. The cavity-molecule coupling is $\bar{g}(t) ={g_0} \exp[{-{(\bar{Y}_0+\mathrm{v}t)^2}/{w^2_0}}]/2$ with the waist of the mode $w_0\simeq11.59$\,$\mathrm{mm}$ and $g_0=2\pi\times 3.68$\,$\mathrm{Hz}$. $\bar{Y}_0$ is the initial position of the sample. $\mathrm{v}$ is the forward velocity of the sample. The detuning is chosen as $\Delta_m=2\pi\times822.7$\,$\mathrm{Hz}$. 
More details about the molecular and cavity parameters, as well as the design of our working cavity, are shown in Sec.\,1.2 of Supporting Information.

The initial photon number is $N_0=\eta^2/\kappa^2$. By adjusting the pumping rate $\eta$, we can tune the hybrid system from the dispersive region with $\lambda\equiv N_0/N_{\mathrm{cr}}\ll 1$ to the non-dispersive region $\lambda > 1$ with the critical photon number of the dispersive region {$N_{\mathrm{cr}}=4|\Delta_m|/g^2_0$}. In Fig.\,\ref{Fig2}, we show the numerical results of $\mathcal{N}$ and $\sigma^z$ for enantiopure samples of left- (blue dashed lines) and right-handed (red solid lines) molecules by numerical solving Eq.\,(\ref{SS1}). The corresponding particle number and forward velocity are set as $N_m=1000$ and $\mathrm{v}=1$\,$\mathrm{m/s}$. 

The numerical simulations show that $\mathcal{N}$ change signs for the two hypotheses in both the typical dispersive region with $\lambda=0.01$ [see Fig.\,\ref{Fig2}(a)] and the typical non-dispersive region with $\lambda=100$ [see Fig.\,\ref{Fig2}(c)]. This clearly demonstrated the ability of our scheme in distinguishing molecular chirality. The time evolutions of the population difference ($\sigma^z$) in the two cases are given in Fig.\,\ref{Fig2}(b) and Fig.\,\ref{Fig2}(d), which clearly show the non-destructive feature of our scheme. In the dispersive case [see Fig.\,\ref{Fig2}(b)], $\sigma^z$ is almost unchanged in the whole process. In the non-dispersive case, $\sigma^z$ is changed when the sample is inside the cavity [see $2<t/\tau<6$ in Fig.\,\ref{Fig2}(d)]. The time unit is $\tau\equiv w_0/\mathrm{v}$. Yet, it almost returns to the initial value [see $t/\tau>6$ in Fig.\,\ref{Fig2}(d)]. Therefore, the molecular state remains unchanged after the detection in both two cases, i.e., the non-destructive detection.   

To evaluate the efficiency of our scheme, we turn to the single-shot signal-to-noise rate $\mathrm{SNR}\equiv{|\bar{n}_L-\bar{n}_R|}/{(2\delta)}$, which gives the error probability as $P_{\mathrm{err}}=\mathrm{Erfc}(\mathrm{SNR}/\sqrt{2})/2$. Here, $\mathrm{Erfc}(x)$ is the complementary error function. When $\mathrm{SNR}\ge3$, the error probability is $P_{\mathrm{err}}\le0.001$, i.e., highly credible chiral discrimination via single-shot measurement can be obtained. In our scheme, the signals are collected from {$t_0=-\bar{Y}_0/\mathrm{v}-M_Y\tau$ to $t_f=-\bar{Y}_0/\mathrm{v}+M_Y\tau$ with $\bar{Y}_0/\mathrm{v}=-4\tau$.} 
In the dispersive region, the signal-to-noise rates for the molecular samples moving sufficiently slowly are
\begin{align}\label{SNR2}
    \mathrm{SNR}\simeq \frac{\sqrt{{\kappa N_0} w_{0}\pi}}{4\sqrt{2\mathrm{v}M_Y}}\mathrm{Erf}(\sqrt{2}M_{Y})\frac{ g^2_0 N_m}{\kappa |\Delta_m| }.
\end{align}
In the dispersive region, the optimized signal-to-noise rates with respect to $M_Y$ are obtained at $M_{Y}\simeq 0.7$. Therefore, our signals are collected from $t_0\simeq3.3\tau$ to $t_f\simeq4.7\tau$. For the deduction of Eq.\,(\ref{SNR2}) in the dispersive limit, see Sec.\,3.4 of Supporting Information. 

We are interested in how the signal-to-noise rates vary with the variations of $N_m$, $\lambda$, and $\mathrm{v}$. To this end, we numerically solve the equations of motion (\ref{SS1}), then use the numerical results to give the enantioselective mean values $\bar{n}_Q$ according to Eq.\,(\ref{EnQ}), and finally give the signal-to-noise rates according to its definition. The $\mathrm{SNR}$ are linear functions of $N_m$ for different $\lambda$ and $\mathrm{v}$ [see Fig.\,\ref{Fig2}(e) and Fig.\,\ref{Fig2}(f)]. The dashed horizontal lines therein indicate $\mathrm{SNR}=3$, whose intersections with the lines give the critical values of $N_m$ (i.e., $N^{\mathrm{cr}}_m$) for highly credible distinguishing of molecular chirality. We can see that 
for $\mathrm{v}=1$\,$\mathrm{m/s}$, the molecular chirality of samples with $N_{m}\ge N^{\mathrm{cr}}_{m}\simeq 552$ can be distinguished with high credibility in the typical dispersive case with $\lambda=0.01$. With the increase of $\lambda$, our scheme will become more efficient, see the typical non-dispersive case ($\lambda=100$) with $N^{\mathrm{cr}}_{m}\simeq 95$. In Fig.\,\ref{Fig2}(g), we show $N^{\mathrm{cr}}_{m}$ as functions of $\lambda$ at different $\mathrm{v}$. In each case, $N^{\mathrm{cr}}_{m}$ approaches a lower limit with the increase of $\lambda$. Moreover, we show $N^{\mathrm{cr}}_{m}$ in our interested area on the $\lambda-\mathrm{v}$ plane [see Fig.\,\ref{Fig2}(h)], which indicates that our scheme can highly credibly distinguish the molecular chirality of slowly moving enantiopure samples with particle numbers of the order of $10^2\sim10^3$.

\textit{Feasibility.}
In our numerical simulations, we have assumed that the free space decay, the collective decay, and the dipole-dipole interaction between molecules are negligible. In our example of 1,2-propanediol, the free space decay rates for the three transitions are 
$\Gamma_0(2\rightarrow1)=2\pi\times1.8\times10^{-10}$\,Hz, $\Gamma_0(3\rightarrow1)=2\pi\times 3.64\times10^{-11}$\,Hz, and  $\Gamma_0(3\rightarrow2)=2\pi\times 8.06\times10^{-11}$\,Hz. 
They are many orders of magnitude smaller than other parameters ($g_0$, $\kappa$, and $\Delta_m$). The particle numbers of the samples under consideration are small ($N_m\le3000$). The typical geometry of the samples under consideration is about $0.1 w_0\simeq 1$\,mm.  Under these parameters, it is sound to assume that the free space decay and the molecular dipole-dipole interaction are negligible. More details about this claim are shown in Sec.\,2.2 of Supporting Information.} Beyond these, we have also proven that the collective decay is negligible in our discussions by comparing the results of Eq.\,(\ref{SS1}) with the results of the second-order mean-field theory in cumulant expansion that includes collective decay (see Sec.\,2.3 of Supporting Information). Thus, in our issue, the dynamics of the system can be well described by the equation of motion~(\ref{SS1}).

 The two core techniques of our proposal have already been experimentally demonstrated. The first core technique, i.e., the microwave enantio-specific state transfer by manipulating the purely rotational transitions, has been achieved with 1,2-propanediol~\cite{eibenberger2017enantiomer},  menthone~\cite{perez2017coherent},  carvone~\cite{perez2017coherent}, and 1-indanol~\cite{lee2022quantitative}. The second core technique, i.e., the dispersive-detection technique with the help of microwave resonators, has been well established in the state-selective detection of single atoms~\cite{gehr2010cavity,volz2011measurement,hattermann2017coupling,garcia2019single}. Its working band is overlapped with the typical rotational transition frequency of chiral molecules and thus it is promising to use such a technique to detect the molecular rotational states. The preparation of slowly moving samples that initially occupy the ground rotational state is envisioned to be handled with sub-millikelvin cooling and slowing techniques of polyatomic molecules~\cite{lu2014magnetic,glockner2015rotational,prehn2016optoelectrical,kozyryev2017sisyphus,mitra2020direct,augenbraun2020molecular,prehn2021high}. Chiral molecules initially populating in the one rotational state of the cyclic three-level model can also be efficiently prepared at cold temperature ($\sim 1$\,K)  by depleting the original thermal populations in the excited states of the working model~\cite{lee2022quantitative}. The forward velocity of samples can be controlled by using the velocity selection via the Doppler effect~\cite{zhou:tel-00737657}.

\textit{Conclusion.}
We have proposed a quantum sensor for detecting molecular chirality by combining the microwave enantio-specific state transfer and the dispersive-detection technique with the help of a microwave resonator. 
It is promising to distinguish slowly moving ($1\sim10$\,m/s) enantiopure samples of opposite chirality with particle numbers of the order of $10^2\sim10^3$ in a single-shot decision. In our scheme, chiral molecules return to their enantioselective final states of enantio-specific state transfer after the whole detection process, thus are not destroyed or disturbed after detection. We note that chiral discrimination at the single-molecule level can be obtained in our scheme by further trapping chiral molecules in the cavity for more than $100$\,s (for more details, see Sec.\,4 of Supporting Information). It is experimentally available to trap polyatomic molecules in an electrostatic trap with typical geometry of $3$\,mm for up to a minute~\cite{prehn2016optoelectrical}. 
For chiral molecules, promising trapping techniques are under development~\cite{augenbraun2020molecular}. 

Beyond this, we have for the first time established the connection between chiral discrimination and quantum hypothesis testing. To our knowledge, there are other state-selective techniques with the ability in quantum hypothesis testing based on physical phenomena including electromagnetically induced transparency~\cite{gunter2013observing}, optical birefringence~\cite{guan2020nondestructive}, F\"{o}rster resonant energy transfer~\cite{jarisch2018state}, and frequency modulation 
spectroscopy~\cite{savalli1999optical}. In this sense, our work will not only stimulate experimentalists using our scheme for chiral discrimination but also trigger the application of other promising non-destructive detection techniques in the very prospective research field.

\begin{acknowledgement}
The authors thank the Supporting by
National Natural Science Foundation of China (12105011, 91850205, 12074030, and 11904022).
\end{acknowledgement}

\begin{suppinfo}
{Additional details on the physical implementation of our scheme, the mean-field theory of our driven Tavis-Cummings model, the balanced Homodyne detection in the single-shot decision of molecular chirality, and the chiral discrimination of trapped single molecule.}
\end{suppinfo}

\bibliography{achemso-demo}

\providecommand{\latin}[1]{#1}
\makeatletter
\providecommand{\doi}
  {\begingroup\let\do\@makeother\dospecials
  \catcode`\{=1 \catcode`\}=2 \doi@aux}
\providecommand{\doi@aux}[1]{\endgroup\texttt{#1}}
\makeatother
\providecommand*\mcitethebibliography{\thebibliography}
\csname @ifundefined\endcsname{endmcitethebibliography}
  {\let\endmcitethebibliography\endthebibliography}{}
\begin{mcitethebibliography}{46}
\providecommand*\natexlab[1]{#1}
\providecommand*\mciteSetBstSublistMode[1]{}
\providecommand*\mciteSetBstMaxWidthForm[2]{}
\providecommand*\mciteBstWouldAddEndPuncttrue
  {\def\EndOfBibitem{\unskip.}}
\providecommand*\mciteBstWouldAddEndPunctfalse
  {\let\EndOfBibitem\relax}
\providecommand*\mciteSetBstMidEndSepPunct[3]{}
\providecommand*\mciteSetBstSublistLabelBeginEnd[3]{}
\providecommand*\EndOfBibitem{}
\mciteSetBstSublistMode{f}
\mciteSetBstMaxWidthForm{subitem}{(\alph{mcitesubitemcount})}
\mciteSetBstSublistLabelBeginEnd
  {\mcitemaxwidthsubitemform\space}
  {\relax}
  {\relax}

\bibitem[Pasteur(1915)]{pasteur1915researches}
Pasteur,~L. \emph{Researches on the molecular asymmetry of natural organic
  products}; University of Chicago Press, 1915\relax
\mciteBstWouldAddEndPuncttrue
\mciteSetBstMidEndSepPunct{\mcitedefaultmidpunct}
{\mcitedefaultendpunct}{\mcitedefaultseppunct}\relax
\EndOfBibitem
\bibitem[Salam and Meath(1997)Salam, and Meath]{salam1997control}
Salam,~A.; Meath,~W. On the control of excited state relative populations of
  enantiomers using circularly polarized pulses of varying durations. \emph{The
  Journal of Chemical Physics} \textbf{1997}, \emph{106}, 7865--7868\relax
\mciteBstWouldAddEndPuncttrue
\mciteSetBstMidEndSepPunct{\mcitedefaultmidpunct}
{\mcitedefaultendpunct}{\mcitedefaultseppunct}\relax
\EndOfBibitem
\bibitem[Shao and H{\"a}nggi(1997)Shao, and H{\"a}nggi]{shao1997control}
Shao,~J.; H{\"a}nggi,~P. Control of molecular chirality. \emph{The Journal of
  Chemical Physics} \textbf{1997}, \emph{107}, 9935--9941\relax
\mciteBstWouldAddEndPuncttrue
\mciteSetBstMidEndSepPunct{\mcitedefaultmidpunct}
{\mcitedefaultendpunct}{\mcitedefaultseppunct}\relax
\EndOfBibitem
\bibitem[Tierney \latin{et~al.}(2011)Tierney, Murphy, and
  Sykes]{tierney2011regular}
Tierney,~H.~L.; Murphy,~C.~J.; Sykes,~E. C.~H. Regular scanning tunneling
  microscope tips can be intrinsically chiral. \emph{Physical Review Letters}
  \textbf{2011}, \emph{106}, 010801\relax
\mciteBstWouldAddEndPuncttrue
\mciteSetBstMidEndSepPunct{\mcitedefaultmidpunct}
{\mcitedefaultendpunct}{\mcitedefaultseppunct}\relax
\EndOfBibitem
\bibitem[Balema \latin{et~al.}(2021)Balema, Liu, Wasio, Larson, Patel,
  Deshlahra, and Sykes]{balema2021enantioselective}
Balema,~T.~A.; Liu,~Y.; Wasio,~N.~A.; Larson,~A.~M.; Patel,~D.~A.;
  Deshlahra,~P.; Sykes,~E. C.~H. Enantioselective Effects in the Electrical
  Excitation of Amine Single-Molecule Rotors. \emph{The Journal of Physical
  Chemistry C} \textbf{2021}, \emph{125}, 3584--3589\relax
\mciteBstWouldAddEndPuncttrue
\mciteSetBstMidEndSepPunct{\mcitedefaultmidpunct}
{\mcitedefaultendpunct}{\mcitedefaultseppunct}\relax
\EndOfBibitem
\bibitem[Patterson \latin{et~al.}(2013)Patterson, Schnell, and
  Doyle]{patterson2013enantiomer}
Patterson,~D.; Schnell,~M.; Doyle,~J.~M. Enantiomer-specific detection of
  chiral molecules via microwave spectroscopy. \emph{Nature} \textbf{2013},
  \emph{497}, 475--477\relax
\mciteBstWouldAddEndPuncttrue
\mciteSetBstMidEndSepPunct{\mcitedefaultmidpunct}
{\mcitedefaultendpunct}{\mcitedefaultseppunct}\relax
\EndOfBibitem
\bibitem[Patterson and Doyle(2013)Patterson, and Doyle]{patterson2013sensitive}
Patterson,~D.; Doyle,~J.~M. Sensitive chiral analysis via microwave three-wave
  mixing. \emph{Physical Review Letters} \textbf{2013}, \emph{111},
  023008\relax
\mciteBstWouldAddEndPuncttrue
\mciteSetBstMidEndSepPunct{\mcitedefaultmidpunct}
{\mcitedefaultendpunct}{\mcitedefaultseppunct}\relax
\EndOfBibitem
\bibitem[Patterson and Schnell(2014)Patterson, and Schnell]{patterson2014new}
Patterson,~D.; Schnell,~M. New studies on molecular chirality in the gas phase:
  enantiomer differentiation and determination of enantiomeric excess.
  \emph{Physical Chemistry Chemical Physics} \textbf{2014}, \emph{16},
  11114--11123\relax
\mciteBstWouldAddEndPuncttrue
\mciteSetBstMidEndSepPunct{\mcitedefaultmidpunct}
{\mcitedefaultendpunct}{\mcitedefaultseppunct}\relax
\EndOfBibitem
\bibitem[Shubert \latin{et~al.}(2014)Shubert, Schmitz, Patterson, Doyle, and
  Schnell]{shubert2014identifying}
Shubert,~V.~A.; Schmitz,~D.; Patterson,~D.; Doyle,~J.~M.; Schnell,~M.
  Identifying enantiomers in mixtures of chiral molecules with broadband
  microwave spectroscopy. \emph{Angewandte Chemie International Edition}
  \textbf{2014}, \emph{53}, 1152--1155\relax
\mciteBstWouldAddEndPuncttrue
\mciteSetBstMidEndSepPunct{\mcitedefaultmidpunct}
{\mcitedefaultendpunct}{\mcitedefaultseppunct}\relax
\EndOfBibitem
\bibitem[Shubert \latin{et~al.}(2015)Shubert, Schmitz, Medcraft, Krin,
  Patterson, Doyle, and Schnell]{shubert2015rotational}
Shubert,~V.~A.; Schmitz,~D.; Medcraft,~C.; Krin,~A.; Patterson,~D.;
  Doyle,~J.~M.; Schnell,~M. Rotational spectroscopy and three-wave mixing of
  4-carvomenthenol: A technical guide to measuring chirality in the microwave
  regime. \emph{The Journal of Chemical physics} \textbf{2015}, \emph{142},
  214201\relax
\mciteBstWouldAddEndPuncttrue
\mciteSetBstMidEndSepPunct{\mcitedefaultmidpunct}
{\mcitedefaultendpunct}{\mcitedefaultseppunct}\relax
\EndOfBibitem
\bibitem[Lobsiger \latin{et~al.}(2015)Lobsiger, Perez, Evangelisti, Lehmann,
  and Pate]{lobsiger2015molecular}
Lobsiger,~S.; Perez,~C.; Evangelisti,~L.; Lehmann,~K.~K.; Pate,~B.~H. Molecular
  structure and chirality detection by Fourier transform microwave
  spectroscopy. \emph{The Journal of Physical Chemistry Letters} \textbf{2015},
  \emph{6}, 196--200\relax
\mciteBstWouldAddEndPuncttrue
\mciteSetBstMidEndSepPunct{\mcitedefaultmidpunct}
{\mcitedefaultendpunct}{\mcitedefaultseppunct}\relax
\EndOfBibitem
\bibitem[Shubert \latin{et~al.}(2016)Shubert, Schmitz, P{\'e}rez, Medcraft,
  Krin, Domingos, Patterson, and Schnell]{shubert2016chiral}
Shubert,~V.~A.; Schmitz,~D.; P{\'e}rez,~C.; Medcraft,~C.; Krin,~A.;
  Domingos,~S.~R.; Patterson,~D.; Schnell,~M. Chiral analysis using broadband
  rotational spectroscopy. \emph{The Journal of Physical Chemistry Letters}
  \textbf{2016}, \emph{7}, 341--350\relax
\mciteBstWouldAddEndPuncttrue
\mciteSetBstMidEndSepPunct{\mcitedefaultmidpunct}
{\mcitedefaultendpunct}{\mcitedefaultseppunct}\relax
\EndOfBibitem
\bibitem[Grabow(2013)]{grabow2013fourier}
Grabow,~J.-U. Fourier transform microwave spectroscopy: handedness caught by
  rotational coherence. \emph{Angewandte Chemie International Edition}
  \textbf{2013}, \emph{52}, 11698--11700\relax
\mciteBstWouldAddEndPuncttrue
\mciteSetBstMidEndSepPunct{\mcitedefaultmidpunct}
{\mcitedefaultendpunct}{\mcitedefaultseppunct}\relax
\EndOfBibitem
\bibitem[Eibenberger \latin{et~al.}(2017)Eibenberger, Doyle, and
  Patterson]{eibenberger2017enantiomer}
Eibenberger,~S.; Doyle,~J.; Patterson,~D. Enantiomer-specific state transfer of
  chiral molecules. \emph{Physical Review Letters} \textbf{2017}, \emph{118},
  123002\relax
\mciteBstWouldAddEndPuncttrue
\mciteSetBstMidEndSepPunct{\mcitedefaultmidpunct}
{\mcitedefaultendpunct}{\mcitedefaultseppunct}\relax
\EndOfBibitem
\bibitem[P{\'e}rez \latin{et~al.}(2017)P{\'e}rez, Steber, Domingos, Krin,
  Schmitz, and Schnell]{perez2017coherent}
P{\'e}rez,~C.; Steber,~A.~L.; Domingos,~S.~R.; Krin,~A.; Schmitz,~D.;
  Schnell,~M. Coherent Enantiomer-Selective Population Enrichment Using
  Tailored Microwave Fields. \emph{Angewandte Chemie International Edition}
  \textbf{2017}, \emph{56}, 12512--12517\relax
\mciteBstWouldAddEndPuncttrue
\mciteSetBstMidEndSepPunct{\mcitedefaultmidpunct}
{\mcitedefaultendpunct}{\mcitedefaultseppunct}\relax
\EndOfBibitem
\bibitem[Lee \latin{et~al.}(2022)Lee, Bischoff, Hernandez-Castillo, Sartakov,
  Meijer, and Eibenberger-Arias]{lee2022quantitative}
Lee,~J.; Bischoff,~J.; Hernandez-Castillo,~A.; Sartakov,~B.; Meijer,~G.;
  Eibenberger-Arias,~S. Quantitative study of enantiomer-specific state
  transfer. \emph{Physical Review Letters} \textbf{2022}, \emph{128},
  173001\relax
\mciteBstWouldAddEndPuncttrue
\mciteSetBstMidEndSepPunct{\mcitedefaultmidpunct}
{\mcitedefaultendpunct}{\mcitedefaultseppunct}\relax
\EndOfBibitem
\bibitem[Ye \latin{et~al.}(2018)Ye, Zhang, and Li]{ye2018real}
Ye,~C.; Zhang,~Q.; Li,~Y. Real single-loop cyclic three-level configuration of
  chiral molecules. \emph{Physical Review A} \textbf{2018}, \emph{98},
  063401\relax
\mciteBstWouldAddEndPuncttrue
\mciteSetBstMidEndSepPunct{\mcitedefaultmidpunct}
{\mcitedefaultendpunct}{\mcitedefaultseppunct}\relax
\EndOfBibitem
\bibitem[Kr{\'a}l and Shapiro(2001)Kr{\'a}l, and Shapiro]{kral2001cyclic}
Kr{\'a}l,~P.; Shapiro,~M. Cyclic population transfer in quantum systems with
  broken symmetry. \emph{Physical Review Letters} \textbf{2001}, \emph{87},
  183002\relax
\mciteBstWouldAddEndPuncttrue
\mciteSetBstMidEndSepPunct{\mcitedefaultmidpunct}
{\mcitedefaultendpunct}{\mcitedefaultseppunct}\relax
\EndOfBibitem
\bibitem[Kr{\'a}l \latin{et~al.}(2003)Kr{\'a}l, Thanopulos, Shapiro, and
  Cohen]{kral2003two}
Kr{\'a}l,~P.; Thanopulos,~I.; Shapiro,~M.; Cohen,~D. Two-step enantio-selective
  optical switch. \emph{Physical Review Letters} \textbf{2003}, \emph{90},
  033001\relax
\mciteBstWouldAddEndPuncttrue
\mciteSetBstMidEndSepPunct{\mcitedefaultmidpunct}
{\mcitedefaultendpunct}{\mcitedefaultseppunct}\relax
\EndOfBibitem
\bibitem[Li \latin{et~al.}(2007)Li, Bruder, and Sun]{li2007generalized}
Li,~Y.; Bruder,~C.; Sun,~C. Generalized {S}tern-{G}erlach effect for chiral
  molecules. \emph{Physical Review Letters} \textbf{2007}, \emph{99},
  130403\relax
\mciteBstWouldAddEndPuncttrue
\mciteSetBstMidEndSepPunct{\mcitedefaultmidpunct}
{\mcitedefaultendpunct}{\mcitedefaultseppunct}\relax
\EndOfBibitem
\bibitem[Vitanov and Drewsen(2019)Vitanov, and Drewsen]{vitanov2019highly}
Vitanov,~N.~V.; Drewsen,~M. Highly efficient detection and separation of chiral
  molecules through shortcuts to adiabaticity. \emph{Physical Review Letters}
  \textbf{2019}, \emph{122}, 173202\relax
\mciteBstWouldAddEndPuncttrue
\mciteSetBstMidEndSepPunct{\mcitedefaultmidpunct}
{\mcitedefaultendpunct}{\mcitedefaultseppunct}\relax
\EndOfBibitem
\bibitem[Ye \latin{et~al.}(2021)Ye, Sun, and Zhang]{ye2021entanglement}
Ye,~C.; Sun,~Y.; Zhang,~X. Entanglement-Assisted Quantum Chiral Spectroscopy.
  \emph{The Journal of Physical Chemistry Letters} \textbf{2021}, \emph{12},
  8591--8597\relax
\mciteBstWouldAddEndPuncttrue
\mciteSetBstMidEndSepPunct{\mcitedefaultmidpunct}
{\mcitedefaultendpunct}{\mcitedefaultseppunct}\relax
\EndOfBibitem
\bibitem[Khokhlova \latin{et~al.}(2022)Khokhlova, Pisanty, Patchkovskii,
  Smirnova, and Ivanov]{khokhlova2022enantiosensitive}
Khokhlova,~M.; Pisanty,~E.; Patchkovskii,~S.; Smirnova,~O.; Ivanov,~M.
  Enantiosensitive steering of free-induction decay. \emph{Science Advances}
  \textbf{2022}, \emph{8}, eabq1962\relax
\mciteBstWouldAddEndPuncttrue
\mciteSetBstMidEndSepPunct{\mcitedefaultmidpunct}
{\mcitedefaultendpunct}{\mcitedefaultseppunct}\relax
\EndOfBibitem
\bibitem[Cai \latin{et~al.}(2022)Cai, Ye, Dong, and
  Li]{cai2022enantiodetection}
Cai,~M.~R.; Ye,~C.; Dong,~H.; Li,~Y. Enantiodetection of chiral molecules via
  two-dimensional spectroscopy. \emph{Physical Review Letters} \textbf{2022},
  \emph{129}, 103201\relax
\mciteBstWouldAddEndPuncttrue
\mciteSetBstMidEndSepPunct{\mcitedefaultmidpunct}
{\mcitedefaultendpunct}{\mcitedefaultseppunct}\relax
\EndOfBibitem
\bibitem[Leibscher \latin{et~al.}(2022)Leibscher, Pozzoli, P{\'e}rez, Schnell,
  Sigalotti, Boscain, and Koch]{leibscher2022full}
Leibscher,~M.; Pozzoli,~E.; P{\'e}rez,~C.; Schnell,~M.; Sigalotti,~M.;
  Boscain,~U.; Koch,~C.~P. Full quantum control of enantiomer-selective state
  transfer in chiral molecules despite degeneracy. \emph{Communications
  Physics} \textbf{2022}, \emph{5}, 1--16\relax
\mciteBstWouldAddEndPuncttrue
\mciteSetBstMidEndSepPunct{\mcitedefaultmidpunct}
{\mcitedefaultendpunct}{\mcitedefaultseppunct}\relax
\EndOfBibitem
\bibitem[Quack(2012)]{quack2012molecular}
Quack,~M. \emph{Quantum Systems in Chemistry and Physics: Progress in Methods
  and Applications}; Springer, 2012; pp 47--76\relax
\mciteBstWouldAddEndPuncttrue
\mciteSetBstMidEndSepPunct{\mcitedefaultmidpunct}
{\mcitedefaultendpunct}{\mcitedefaultseppunct}\relax
\EndOfBibitem
\bibitem[Spedalieri and Braunstein(2014)Spedalieri, and
  Braunstein]{spedalieri2014asymmetric}
Spedalieri,~G.; Braunstein,~S.~L. Asymmetric quantum hypothesis testing with
  Gaussian states. \emph{Physical Review A} \textbf{2014}, \emph{90},
  052307\relax
\mciteBstWouldAddEndPuncttrue
\mciteSetBstMidEndSepPunct{\mcitedefaultmidpunct}
{\mcitedefaultendpunct}{\mcitedefaultseppunct}\relax
\EndOfBibitem
\bibitem[Kubo(1962)]{kubo1962generalized}
Kubo,~R. Generalized cumulant expansion method. \emph{Journal of the Physical
  Society of Japan} \textbf{1962}, \emph{17}, 1100--1120\relax
\mciteBstWouldAddEndPuncttrue
\mciteSetBstMidEndSepPunct{\mcitedefaultmidpunct}
{\mcitedefaultendpunct}{\mcitedefaultseppunct}\relax
\EndOfBibitem
\bibitem[Horak \latin{et~al.}(2003)Horak, Klappauf, Haase, Folman,
  Schmiedmayer, Domokos, and Hinds]{horak2003possibility}
Horak,~P.; Klappauf,~B.~G.; Haase,~A.; Folman,~R.; Schmiedmayer,~J.;
  Domokos,~P.; Hinds,~E. Possibility of single-atom detection on a chip.
  \emph{Physical Review A} \textbf{2003}, \emph{67}, 043806\relax
\mciteBstWouldAddEndPuncttrue
\mciteSetBstMidEndSepPunct{\mcitedefaultmidpunct}
{\mcitedefaultendpunct}{\mcitedefaultseppunct}\relax
\EndOfBibitem
\bibitem[Gehr \latin{et~al.}(2010)Gehr, Volz, Dubois, Steinmetz, Colombe, Lev,
  Long, Esteve, and Reichel]{gehr2010cavity}
Gehr,~R.; Volz,~J.; Dubois,~G.; Steinmetz,~T.; Colombe,~Y.; Lev,~B.~L.;
  Long,~R.; Esteve,~J.; Reichel,~J. Cavity-based single atom preparation and
  high-fidelity hyperfine state readout. \emph{Physical Review Letters}
  \textbf{2010}, \emph{104}, 203602\relax
\mciteBstWouldAddEndPuncttrue
\mciteSetBstMidEndSepPunct{\mcitedefaultmidpunct}
{\mcitedefaultendpunct}{\mcitedefaultseppunct}\relax
\EndOfBibitem
\bibitem[Volz \latin{et~al.}(2011)Volz, Gehr, Dubois, Est{\`e}ve, and
  Reichel]{volz2011measurement}
Volz,~J.; Gehr,~R.; Dubois,~G.; Est{\`e}ve,~J.; Reichel,~J. Measurement of the
  internal state of a single atom without energy exchange. \emph{Nature}
  \textbf{2011}, \emph{475}, 210--213\relax
\mciteBstWouldAddEndPuncttrue
\mciteSetBstMidEndSepPunct{\mcitedefaultmidpunct}
{\mcitedefaultendpunct}{\mcitedefaultseppunct}\relax
\EndOfBibitem
\bibitem[Hattermann \latin{et~al.}(2017)Hattermann, Bothner, Ley, Ferdinand,
  Wiedmaier, S{\'a}rk{\'a}ny, Kleiner, Koelle, and
  Fort{\'a}gh]{hattermann2017coupling}
Hattermann,~H.; Bothner,~D.; Ley,~L.; Ferdinand,~B.; Wiedmaier,~D.;
  S{\'a}rk{\'a}ny,~L.; Kleiner,~R.; Koelle,~D.; Fort{\'a}gh,~J. Coupling
  ultracold atoms to a superconducting coplanar waveguide resonator.
  \emph{Nature Communications} \textbf{2017}, \emph{8}, 1--7\relax
\mciteBstWouldAddEndPuncttrue
\mciteSetBstMidEndSepPunct{\mcitedefaultmidpunct}
{\mcitedefaultendpunct}{\mcitedefaultseppunct}\relax
\EndOfBibitem
\bibitem[Garcia \latin{et~al.}(2019)Garcia, Stammeier, Deiglmayr, Merkt, and
  Wallraff]{garcia2019single}
Garcia,~S.; Stammeier,~M.; Deiglmayr,~J.; Merkt,~F.; Wallraff,~A. Single-shot
  nondestructive detection of {R}ydberg-atom ensembles by transmission
  measurement of a microwave cavity. \emph{Physical Review Letters}
  \textbf{2019}, \emph{123}, 193201\relax
\mciteBstWouldAddEndPuncttrue
\mciteSetBstMidEndSepPunct{\mcitedefaultmidpunct}
{\mcitedefaultendpunct}{\mcitedefaultseppunct}\relax
\EndOfBibitem
\bibitem[Lu \latin{et~al.}(2014)Lu, Kozyryev, Hemmerling, Piskorski, and
  Doyle]{lu2014magnetic}
Lu,~H.-I.; Kozyryev,~I.; Hemmerling,~B.; Piskorski,~J.; Doyle,~J.~M. Magnetic
  trapping of molecules via optical loading and magnetic slowing.
  \emph{Physical Review Letters} \textbf{2014}, \emph{112}, 113006\relax
\mciteBstWouldAddEndPuncttrue
\mciteSetBstMidEndSepPunct{\mcitedefaultmidpunct}
{\mcitedefaultendpunct}{\mcitedefaultseppunct}\relax
\EndOfBibitem
\bibitem[Gl{\"o}ckner \latin{et~al.}(2015)Gl{\"o}ckner, Prehn, Englert, Rempe,
  and Zeppenfeld]{glockner2015rotational}
Gl{\"o}ckner,~R.; Prehn,~A.; Englert,~B.~G.; Rempe,~G.; Zeppenfeld,~M.
  Rotational cooling of trapped polyatomic molecules. \emph{Physical Review
  Letters} \textbf{2015}, \emph{115}, 233001\relax
\mciteBstWouldAddEndPuncttrue
\mciteSetBstMidEndSepPunct{\mcitedefaultmidpunct}
{\mcitedefaultendpunct}{\mcitedefaultseppunct}\relax
\EndOfBibitem
\bibitem[Prehn \latin{et~al.}(2016)Prehn, Ibr{\"u}gger, Gl{\"o}ckner, Rempe,
  and Zeppenfeld]{prehn2016optoelectrical}
Prehn,~A.; Ibr{\"u}gger,~M.; Gl{\"o}ckner,~R.; Rempe,~G.; Zeppenfeld,~M.
  Optoelectrical cooling of polar molecules to submillikelvin temperatures.
  \emph{Physical Review Letters} \textbf{2016}, \emph{116}, 063005\relax
\mciteBstWouldAddEndPuncttrue
\mciteSetBstMidEndSepPunct{\mcitedefaultmidpunct}
{\mcitedefaultendpunct}{\mcitedefaultseppunct}\relax
\EndOfBibitem
\bibitem[Kozyryev \latin{et~al.}(2017)Kozyryev, Baum, Matsuda, Augenbraun,
  Anderegg, Sedlack, and Doyle]{kozyryev2017sisyphus}
Kozyryev,~I.; Baum,~L.; Matsuda,~K.; Augenbraun,~B.~L.; Anderegg,~L.;
  Sedlack,~A.~P.; Doyle,~J.~M. Sisyphus laser cooling of a polyatomic molecule.
  \emph{Physical Review Letters} \textbf{2017}, \emph{118}, 173201\relax
\mciteBstWouldAddEndPuncttrue
\mciteSetBstMidEndSepPunct{\mcitedefaultmidpunct}
{\mcitedefaultendpunct}{\mcitedefaultseppunct}\relax
\EndOfBibitem
\bibitem[Mitra \latin{et~al.}(2020)Mitra, Vilas, Hallas, Anderegg, Augenbraun,
  Baum, Miller, Raval, and Doyle]{mitra2020direct}
Mitra,~D.; Vilas,~N.~B.; Hallas,~C.; Anderegg,~L.; Augenbraun,~B.~L.; Baum,~L.;
  Miller,~C.; Raval,~S.; Doyle,~J.~M. Direct laser cooling of a symmetric top
  molecule. \emph{Science} \textbf{2020}, \emph{369}, 1366--1369\relax
\mciteBstWouldAddEndPuncttrue
\mciteSetBstMidEndSepPunct{\mcitedefaultmidpunct}
{\mcitedefaultendpunct}{\mcitedefaultseppunct}\relax
\EndOfBibitem
\bibitem[Augenbraun \latin{et~al.}(2020)Augenbraun, Doyle, Zelevinsky, and
  Kozyryev]{augenbraun2020molecular}
Augenbraun,~B.~L.; Doyle,~J.~M.; Zelevinsky,~T.; Kozyryev,~I. Molecular
  asymmetry and optical cycling: laser cooling asymmetric top molecules.
  \emph{Physical Review X} \textbf{2020}, \emph{10}, 031022\relax
\mciteBstWouldAddEndPuncttrue
\mciteSetBstMidEndSepPunct{\mcitedefaultmidpunct}
{\mcitedefaultendpunct}{\mcitedefaultseppunct}\relax
\EndOfBibitem
\bibitem[Prehn \latin{et~al.}(2021)Prehn, Ibr{\"u}gger, Rempe, and
  Zeppenfeld]{prehn2021high}
Prehn,~A.; Ibr{\"u}gger,~M.; Rempe,~G.; Zeppenfeld,~M. High-Resolution
  ``Magic''-Field Spectroscopy on Trapped Polyatomic Molecules. \emph{Physical
  Review Letters} \textbf{2021}, \emph{127}, 173602\relax
\mciteBstWouldAddEndPuncttrue
\mciteSetBstMidEndSepPunct{\mcitedefaultmidpunct}
{\mcitedefaultendpunct}{\mcitedefaultseppunct}\relax
\EndOfBibitem
\bibitem[Zhou(2012)]{zhou:tel-00737657}
Zhou,~X. {Field locked to Fock state by quantum feedback with single photon
  corrections}. Theses, {Universit{\'e} Pierre et Marie Curie - Paris VI},
  2012\relax
\mciteBstWouldAddEndPuncttrue
\mciteSetBstMidEndSepPunct{\mcitedefaultmidpunct}
{\mcitedefaultendpunct}{\mcitedefaultseppunct}\relax
\EndOfBibitem
\bibitem[G{\"u}nter \latin{et~al.}(2013)G{\"u}nter, Schempp, Robert-de
  Saint-Vincent, Gavryusev, Helmrich, Hofmann, Whitlock, and
  Weidem{\"u}ller]{gunter2013observing}
G{\"u}nter,~G.; Schempp,~H.; Robert-de Saint-Vincent,~M.; Gavryusev,~V.;
  Helmrich,~S.; Hofmann,~C.; Whitlock,~S.; Weidem{\"u}ller,~M. Observing the
  dynamics of dipole-mediated energy transport by interaction-enhanced imaging.
  \emph{Science} \textbf{2013}, \emph{342}, 954--956\relax
\mciteBstWouldAddEndPuncttrue
\mciteSetBstMidEndSepPunct{\mcitedefaultmidpunct}
{\mcitedefaultendpunct}{\mcitedefaultseppunct}\relax
\EndOfBibitem
\bibitem[Guan \latin{et~al.}(2020)Guan, Highman, Meier, Williams, Scarola,
  DeMarco, Kotochigova, and Gadway]{guan2020nondestructive}
Guan,~Q.; Highman,~M.; Meier,~E.~J.; Williams,~G.~R.; Scarola,~V.; DeMarco,~B.;
  Kotochigova,~S.; Gadway,~B. Nondestructive dispersive imaging of rotationally
  excited ultracold molecules. \emph{Physical Chemistry Chemical Physics}
  \textbf{2020}, \emph{22}, 20531--20544\relax
\mciteBstWouldAddEndPuncttrue
\mciteSetBstMidEndSepPunct{\mcitedefaultmidpunct}
{\mcitedefaultendpunct}{\mcitedefaultseppunct}\relax
\EndOfBibitem
\bibitem[Jarisch and Zeppenfeld(2018)Jarisch, and Zeppenfeld]{jarisch2018state}
Jarisch,~F.; Zeppenfeld,~M. State resolved investigation of F{\"o}rster
  resonant energy transfer in collisions between polar molecules and {R}ydberg
  atoms. \emph{New Journal of Physics} \textbf{2018}, \emph{20}, 113044\relax
\mciteBstWouldAddEndPuncttrue
\mciteSetBstMidEndSepPunct{\mcitedefaultmidpunct}
{\mcitedefaultendpunct}{\mcitedefaultseppunct}\relax
\EndOfBibitem
\bibitem[Savalli \latin{et~al.}(1999)Savalli, Horvath, Featonby, Cognet,
  Westbrook, Westbrook, and Aspect]{savalli1999optical}
Savalli,~V.; Horvath,~G. Z.~K.; Featonby,~P.~D.; Cognet,~L.; Westbrook,~N.;
  Westbrook,~C.~I.; Aspect,~A. Optical detection of cold atoms without
  spontaneous emission. \emph{Optics Letters} \textbf{1999}, \emph{24},
  1552--1554\relax
\mciteBstWouldAddEndPuncttrue
\mciteSetBstMidEndSepPunct{\mcitedefaultmidpunct}
{\mcitedefaultendpunct}{\mcitedefaultseppunct}\relax
\EndOfBibitem
\end{mcitethebibliography}


\begin{thebibliography}{11}%
\makeatletter
\providecommand \@ifxundefined [1]{%
 \@ifx{#1\undefined}
}%
\providecommand \@ifnum [1]{%
 \ifnum #1\expandafter \@firstoftwo
 \else \expandafter \@secondoftwo
 \fi
}%
\providecommand \@ifx [1]{%
 \ifx #1\expandafter \@firstoftwo
 \else \expandafter \@secondoftwo
 \fi
}%
\providecommand \natexlab [1]{#1}%
\providecommand \enquote  [1]{``#1''}%
\providecommand \bibnamefont  [1]{#1}%
\providecommand \bibfnamefont [1]{#1}%
\providecommand \citenamefont [1]{#1}%
\providecommand \href@noop [0]{\@secondoftwo}%
\providecommand \href [0]{\begingroup \@sanitize@url \@href}%
\providecommand \@href[1]{\@@startlink{#1}\@@href}%
\providecommand \@@href[1]{\endgroup#1\@@endlink}%
\providecommand \@sanitize@url [0]{\catcode `\\12\catcode `\$12\catcode
  `\&12\catcode `\#12\catcode `\^12\catcode `\_12\catcode `\%12\relax}%
\providecommand \@@startlink[1]{}%
\providecommand \@@endlink[0]{}%
\providecommand \url  [0]{\begingroup\@sanitize@url \@url }%
\providecommand \@url [1]{\endgroup\@href {#1}{\urlprefix }}%
\providecommand \urlprefix  [0]{URL }%
\providecommand \Eprint [0]{\href }%
\providecommand \doibase [0]{https://doi.org/}%
\providecommand \selectlanguage [0]{\@gobble}%
\providecommand \bibinfo  [0]{\@secondoftwo}%
\providecommand \bibfield  [0]{\@secondoftwo}%
\providecommand \translation [1]{[#1]}%
\providecommand \BibitemOpen [0]{}%
\providecommand \bibitemStop [0]{}%
\providecommand \bibitemNoStop [0]{.\EOS\space}%
\providecommand \EOS [0]{\spacefactor3000\relax}%
\providecommand \BibitemShut  [1]{\csname bibitem#1\endcsname}%
\let\auto@bib@innerbib\@empty
\bibitem [{\citenamefont {Zare}(1988)}]{richard1988angular}%
  \BibitemOpen
  \bibfield  {author} {\bibinfo {author} {\bibfnamefont {R.~N.}\ \bibnamefont
  {Zare}},\ }\href@noop {} {\emph {\bibinfo {title} {Angular momentum:
  understanding spatial aspects in chemistry and physics}}}\ (\bibinfo
  {publisher} {J. Wiley \& Sons},\ \bibinfo {year} {1988})\BibitemShut
  {NoStop}%
\bibitem [{\citenamefont {Ye}\ \emph {et~al.}(2018)\citenamefont {Ye},
  \citenamefont {Zhang},\ and\ \citenamefont {Li}}]{ye2018real}%
  \BibitemOpen
  \bibfield  {author} {\bibinfo {author} {\bibfnamefont {C.}~\bibnamefont
  {Ye}}, \bibinfo {author} {\bibfnamefont {Q.}~\bibnamefont {Zhang}},\ and\
  \bibinfo {author} {\bibfnamefont {Y.}~\bibnamefont {Li}},\ }\bibfield
  {title} {\bibinfo {title} {Real single-loop cyclic three-level configuration
  of chiral molecules},\ }\href@noop {} {\bibfield  {journal} {\bibinfo
  {journal} {Physical Review A}\ }\textbf {\bibinfo {volume} {98}},\ \bibinfo
  {pages} {063401} (\bibinfo {year} {2018})}\BibitemShut {NoStop}%
\bibitem [{\citenamefont {Patterson}\ \emph {et~al.}(2013)\citenamefont
  {Patterson}, \citenamefont {Schnell},\ and\ \citenamefont
  {Doyle}}]{patterson2013enantiomer}%
  \BibitemOpen
  \bibfield  {author} {\bibinfo {author} {\bibfnamefont {D.}~\bibnamefont
  {Patterson}}, \bibinfo {author} {\bibfnamefont {M.}~\bibnamefont {Schnell}},\
  and\ \bibinfo {author} {\bibfnamefont {J.~M.}\ \bibnamefont {Doyle}},\
  }\bibfield  {title} {\bibinfo {title} {Enantiomer-specific detection of
  chiral molecules via microwave spectroscopy},\ }\href@noop {} {\bibfield
  {journal} {\bibinfo  {journal} {Nature}\ }\textbf {\bibinfo {volume} {497}},\
  \bibinfo {pages} {475} (\bibinfo {year} {2013})}\BibitemShut {NoStop}%
\bibitem [{\citenamefont {Zhou}(2012)}]{zhou:tel-00737657}%
  \BibitemOpen
  \bibfield  {author} {\bibinfo {author} {\bibfnamefont {X.}~\bibnamefont
  {Zhou}},\ }\emph {\bibinfo {title} {{Field locked to Fock state by quantum
  feedback with single photon corrections}}},\ \href
  {https://tel.archives-ouvertes.fr/tel-00737657} {\bibinfo {type} {Theses}},\
  \bibinfo  {school} {{Universit{\'e} Pierre et Marie Curie - Paris VI}}
  (\bibinfo {year} {2012})\BibitemShut {NoStop}%
\bibitem [{\citenamefont {Lehmberg}(1970)}]{lehmberg1970radiation}%
  \BibitemOpen
  \bibfield  {author} {\bibinfo {author} {\bibfnamefont {R.}~\bibnamefont
  {Lehmberg}},\ }\bibfield  {title} {\bibinfo {title} {Radiation from an n-atom
  system. i. general formalism},\ }\href@noop {} {\bibfield  {journal}
  {\bibinfo  {journal} {Physical Review A}\ }\textbf {\bibinfo {volume} {2}},\
  \bibinfo {pages} {883} (\bibinfo {year} {1970})}\BibitemShut {NoStop}%
\bibitem [{\citenamefont {Plankensteiner}\ \emph {et~al.}(2017)\citenamefont
  {Plankensteiner}, \citenamefont {Sommer}, \citenamefont {Ritsch},\ and\
  \citenamefont {Genes}}]{plankensteiner2017cavity}%
  \BibitemOpen
  \bibfield  {author} {\bibinfo {author} {\bibfnamefont {D.}~\bibnamefont
  {Plankensteiner}}, \bibinfo {author} {\bibfnamefont {C.}~\bibnamefont
  {Sommer}}, \bibinfo {author} {\bibfnamefont {H.}~\bibnamefont {Ritsch}},\
  and\ \bibinfo {author} {\bibfnamefont {C.}~\bibnamefont {Genes}},\ }\bibfield
   {title} {\bibinfo {title} {Cavity antiresonance spectroscopy of dipole
  coupled subradiant arrays},\ }\href@noop {} {\bibfield  {journal} {\bibinfo
  {journal} {Physical Review Letters}\ }\textbf {\bibinfo {volume} {119}},\
  \bibinfo {pages} {093601} (\bibinfo {year} {2017})}\BibitemShut {NoStop}%
\bibitem [{\citenamefont {Plankensteiner}\ \emph {et~al.}(2019)\citenamefont
  {Plankensteiner}, \citenamefont {Sommer}, \citenamefont {Reitz},
  \citenamefont {Ritsch},\ and\ \citenamefont
  {Genes}}]{plankensteiner2019enhanced}%
  \BibitemOpen
  \bibfield  {author} {\bibinfo {author} {\bibfnamefont {D.}~\bibnamefont
  {Plankensteiner}}, \bibinfo {author} {\bibfnamefont {C.}~\bibnamefont
  {Sommer}}, \bibinfo {author} {\bibfnamefont {M.}~\bibnamefont {Reitz}},
  \bibinfo {author} {\bibfnamefont {H.}~\bibnamefont {Ritsch}},\ and\ \bibinfo
  {author} {\bibfnamefont {C.}~\bibnamefont {Genes}},\ }\bibfield  {title}
  {\bibinfo {title} {Enhanced collective purcell effect of coupled quantum
  emitter systems},\ }\href@noop {} {\bibfield  {journal} {\bibinfo  {journal}
  {Physical Review A}\ }\textbf {\bibinfo {volume} {99}},\ \bibinfo {pages}
  {043843} (\bibinfo {year} {2019})}\BibitemShut {NoStop}%
\bibitem [{\citenamefont {Chen}\ \emph {et~al.}(2021)\citenamefont {Chen},
  \citenamefont {Ye},\ and\ \citenamefont {Li}}]{chen2021enantio}%
  \BibitemOpen
  \bibfield  {author} {\bibinfo {author} {\bibfnamefont {Y.-Y.}\ \bibnamefont
  {Chen}}, \bibinfo {author} {\bibfnamefont {C.}~\bibnamefont {Ye}},\ and\
  \bibinfo {author} {\bibfnamefont {Y.}~\bibnamefont {Li}},\ }\bibfield
  {title} {\bibinfo {title} {Enantio-detection via cavity-assisted three-photon
  processes},\ }\href@noop {} {\bibfield  {journal} {\bibinfo  {journal}
  {Optics Express}\ }\textbf {\bibinfo {volume} {29}},\ \bibinfo {pages}
  {36132} (\bibinfo {year} {2021})}\BibitemShut {NoStop}%
\bibitem [{\citenamefont {Kubo}(1962)}]{kubo1962generalized}%
  \BibitemOpen
  \bibfield  {author} {\bibinfo {author} {\bibfnamefont {R.}~\bibnamefont
  {Kubo}},\ }\bibfield  {title} {\bibinfo {title} {Generalized cumulant
  expansion method},\ }\href@noop {} {\bibfield  {journal} {\bibinfo  {journal}
  {Journal of the Physical Society of Japan}\ }\textbf {\bibinfo {volume}
  {17}},\ \bibinfo {pages} {1100} (\bibinfo {year} {1962})}\BibitemShut
  {NoStop}%
\bibitem [{\citenamefont {Horak}\ \emph {et~al.}(2003)\citenamefont {Horak},
  \citenamefont {Klappauf}, \citenamefont {Haase}, \citenamefont {Folman},
  \citenamefont {Schmiedmayer}, \citenamefont {Domokos},\ and\ \citenamefont
  {Hinds}}]{horak2003possibility}%
  \BibitemOpen
  \bibfield  {author} {\bibinfo {author} {\bibfnamefont {P.}~\bibnamefont
  {Horak}}, \bibinfo {author} {\bibfnamefont {B.~G.}\ \bibnamefont {Klappauf}},
  \bibinfo {author} {\bibfnamefont {A.}~\bibnamefont {Haase}}, \bibinfo
  {author} {\bibfnamefont {R.}~\bibnamefont {Folman}}, \bibinfo {author}
  {\bibfnamefont {J.}~\bibnamefont {Schmiedmayer}}, \bibinfo {author}
  {\bibfnamefont {P.}~\bibnamefont {Domokos}},\ and\ \bibinfo {author}
  {\bibfnamefont {E.}~\bibnamefont {Hinds}},\ }\bibfield  {title} {\bibinfo
  {title} {Possibility of single-atom detection on a chip},\ }\href@noop {}
  {\bibfield  {journal} {\bibinfo  {journal} {Physical Review A}\ }\textbf
  {\bibinfo {volume} {67}},\ \bibinfo {pages} {043806} (\bibinfo {year}
  {2003})}\BibitemShut {NoStop}%
\bibitem [{\citenamefont {Barzanjeh}\ \emph {et~al.}(2015)\citenamefont
  {Barzanjeh}, \citenamefont {Guha}, \citenamefont {Weedbrook}, \citenamefont
  {Vitali}, \citenamefont {Shapiro},\ and\ \citenamefont
  {Pirandola}}]{barzanjeh2015microwave}%
  \BibitemOpen
  \bibfield  {author} {\bibinfo {author} {\bibfnamefont {S.}~\bibnamefont
  {Barzanjeh}}, \bibinfo {author} {\bibfnamefont {S.}~\bibnamefont {Guha}},
  \bibinfo {author} {\bibfnamefont {C.}~\bibnamefont {Weedbrook}}, \bibinfo
  {author} {\bibfnamefont {D.}~\bibnamefont {Vitali}}, \bibinfo {author}
  {\bibfnamefont {J.~H.}\ \bibnamefont {Shapiro}},\ and\ \bibinfo {author}
  {\bibfnamefont {S.}~\bibnamefont {Pirandola}},\ }\bibfield  {title} {\bibinfo
  {title} {Microwave quantum illumination},\ }\href@noop {} {\bibfield
  {journal} {\bibinfo  {journal} {Physical Review Letters}\ }\textbf {\bibinfo
  {volume} {114}},\ \bibinfo {pages} {080503} (\bibinfo {year}
  {2015})}\BibitemShut {NoStop}%
\end{thebibliography}%

\end{document}


\preprint{APS/123-QED}
\title{Supplementary materials for ``Single-shot Non-destructive Quantum Sensing
for Gaseous Samples with Hundreds of Chiral
Molecules"}

\author{Chong Ye}
\affiliation{Beijing Key Laboratory of Nanophotonics and Ultrafine Optoelectronic Systems, School of Physics, Beijing Institute of Technology, 100081 Beijing, China}
\author{Yifan Sun}
\affiliation{Beijing Key Laboratory of Nanophotonics and Ultrafine Optoelectronic Systems, School of Physics, Beijing Institute of Technology, 100081 Beijing, China}
\author{Yong Li}\email{yongli@hainanu.edu.cn}
\affiliation{Center for Theoretical Physics and School of Science, Hainan University, Haikou 570228, China}
\affiliation{Synergetic Innovation Center for Quantum Effects and Applications,
Hunan Normal University, Changsha 410081, China}
\author{Xiangdong Zhang}\email{zhangxd@bit.edu.cn}
\affiliation{Beijing Key Laboratory of Nanophotonics and Ultrafine Optoelectronic Systems, School of Physics, Beijing Institute of Technology, 100081 Beijing, China}

\date{\today}
\maketitle


\tableofcontents
\newpage

\section{Physical implementation of our scheme}
\subsection{Enantio-specific state transfer in the cyclic three-level model of chiral molecules}
\subsubsection{Rotational Hamiltonian and electric dipole transition selection rules}
The rotations of chiral molecules are described as the asymmetric top Hamiltonian~\cite{richard1988angular}
\begin{align}
    H_{\mathrm{rot}}=\hbar (A \hat{J}^2_a+B \hat{J}^2_b+C\hat{J}^2_c)
\end{align}
with the rotational constants of the three principal axes $A>B>C$. The angular momentum operators $\hat{J}_a$, $\hat{J}_b$, and 
$\hat{J}_c$ corresponds to the components of the total angular momentum with respect to the principal axes. 
The eigenstates $|J_{K_a,K_c,M}\rangle$ are magnetic degenerated, where $M$ are the magnetic quantum number. The energy levels are designated by $J_{K_a,K_c}$. $J$ is the quantum number of the total angular momentum. $K_a$ and $K_c$ are not quantum numbers. $K_a$ runs from $0$ to $J$. $K_c$ runs from $J$ to $0$. The possible energy levels satisfy $K_a+K_c=J$ or $K_a+K_c=J+1$, such as $J_{J,0}$, $J_{J,1}$, $J_{J-1,1}$, $J_{J-1,2}$. The eigenstates can also be described in the 
$|J_{\tau,M}\rangle$ notation with $\tau=K_a-K_c$ ranging from $-J$ to $J$~\cite{richard1988angular}.

The electric dipole transition selection rules of $J$ are $\Delta J=0,\pm1$~\cite{richard1988angular}. 
The electric dipole transition selection rules of $m$ are $\Delta m=0, \pm1$~\cite{richard1988angular}. The transition dipole for $\Delta m=\sigma$ ($\sigma=0,\pm1$) is in the $\bm\varepsilon_\sigma$ direction with $\bm\varepsilon_0=\bm{Z}$, and 
$\bm\varepsilon_{\pm1}=(\pm\bm{X}+\mathrm{i}\bm{Y})/\sqrt{2}$. Here, $\bm{X}$, $\bm{Y}$, and $\bm{Z}$ are the direction vectors of the three coordinates in the space-fixed frame. The principal axes $a$, $b$, and $c$ make up a coordinate system in the molecule-fixed frame. 
When the molecule rotates, the molecule-fixed frame changes correspondingly and the space-fixed frame is unchanged. 
With respect to the molecule-fixed frame, the electric-dipole-transition-selection-rule allowed transitions are classified to $a$-type, $b$-type, and $c$-type~\cite{richard1988angular}. For $a$-type transitions, $\Delta K_a$ are an even integer, $\Delta K_c$ are an odd integer, and the transition dipoles are proportional to 
$\mu^{Q}_a$  the component of $\langle v^{Q}_g|\bm{\hat{\mu}} |v^{Q}_g\rangle$ in the $a$-axis of the molecule-fixed frame. Here, $\bm{\hat{\mu}}$ is the electric dipole operator~\cite{richard1988angular}. 
For $b$-type transitions, $\Delta K_a$ are an odd integer, $\Delta K_c$ are an odd integer, and the transition dipoles are proportional to 
$\mu^{Q}_b$  the component of $\langle v^{Q}_g|\bm{\hat{\mu}} |v^{Q}_g\rangle$ in the $b$-axis of the molecule-fixed frame~\cite{richard1988angular}. For $c$-type transitions, $\Delta K_a$ are an odd integer, $\Delta K_c$ are an even integer, and the transition dipoles are proportional to 
$\mu^{Q}_c$  the component of $\langle v^{Q}_g|\bm{\hat{\mu}} |v^{Q}_g\rangle$ in the $c$-axis of the molecule-fixed frame~\cite{richard1988angular}. 

\subsubsection{Construct single-loop three-level cyclic model of chiral molecules}
According to the selection rules, we choose a single-loop three-level cyclic configuration of the chiral molecule composed of  
three working states~\cite{ye2018real} $|1\rangle=|v^{Q}_g\rangle|0_{0,0,0}\rangle$, $|2\rangle=|v^{Q}_g\rangle|1_{0,1,0}\rangle$, and $|3\rangle=|v^{Q}_g\rangle(|1_{1,0,1}\rangle+|1_{1,0,-1})/\sqrt{2}$.  $|v^{Q}_g\rangle$ is the chiral vibrational ground state of the energy potential surface of the ground electronic state. The transition angular frequencies are $\omega_{21}=B+C$, $\omega_{31}=A+B$, and 
$\omega_{32}=A-C$. Three phase-controlled classical electromagnetic fields $\bm{E}_{21}$, $\bm{E}_{31}$, and $\bm{E}_{32}$ are applied to couple the single-loop three-level cyclic configuration to prepare the most distinguishable enantiomeric quantum hypotheses. Their frequencies and polarizations should be specified. In our discussions, the fields $E_{ji}$  ($j>i$) are resonantly coupled with the transitions $|i\rangle\leftrightarrow|j\rangle$. $\bm{E}_{21}$, $\bm{E}_{31}$, and $\bm{E}_{32}$ are $Z$-polarized, $Y$-polarized, and $X$-polarized, respectively. In the interaction picture, the corresponding Hamiltonian is
\begin{align}
    \hat{H}_{Q}=\sum^{3}_{j>i=1}(\Omega^{Q}_{ji}|j\rangle\langle i|+h.c.),
\end{align}
where the coupling strengths are $\Omega^{Q}_{21}=E_{21}\mu^{Q}_a e^{\mathrm{i}\phi_{21}}/2$, $\Omega_{31}=E_{31}\mu^{Q}_c e^{\mathrm{i}\phi_{31}}/2\sqrt{3}$, and $\Omega_{32}=E_{32}\mu^{Q}_b e^{\mathrm{i}\phi_{32}}/4$ with $E_{ji}$ being the amplitude of each fields~\cite{ye2018real}. For chiral molecules, we have 
\begin{align}
    |\mu^{L}_{\beta}|=|\mu^{R}_{\beta}|,~~
    \mu^{L}_a\mu^{L}_b\mu^{L}_c=-\mu^{R}_a\mu^{R}_b\mu^{R}_c
\end{align}
with $\beta=a,b,c$. By choosing an appropriate gauge, we arrive 
\begin{align}
    \Omega^{Q}_{21}=\Omega_{21}\equiv E_{21}|\mu^{Q}_a|/2,~~   \Omega^{Q}_{32}=\Omega_{32}\equiv E_{32}|\mu^{Q}_b|/4,~~ 
     \Omega^{L}_{31}=-\Omega^{R}_{31}=\Omega_{31}e^{\mathrm{i}\phi}\equiv E_{31}|\mu^{Q}_c|e^{\mathrm{i}\phi}/2\sqrt{3},
\end{align}
where the overall phase $\phi$ is modulated by adjusting the relative phases of the three applied fields. 

\subsubsection{Non-overlapped resonance pulses}
For enantio-specific state transfer, we apply the $|2\rangle\xleftarrow{\pi/2}|1\rangle\xrightarrow{\pi}|3\rangle\xrightarrow{\pi/2}|2\rangle$ pulses, where the $\pi/2$ pulse creates a $50:50$ superposition between two levels and the $\pi$ pulse inverts the population of two states. Since there is no overlapping for the three pulses, the process can be treated as a series of two-level transitions. We assume that the initial state of the chiral molecule is $|1\rangle$. The final states of the two enantiomers after the first pulse of $|2\rangle\xleftarrow{\pi/2}|1\rangle$ are 
\begin{align}
    |\Psi_{\mathrm{I},Q}\rangle=\frac{1}{\sqrt{2}}(|1\rangle+\mathrm{i}|2\rangle)
\end{align}
with the corresponding $\phi_{\mathrm{TL}}=0$. The finial states of the two enantiomers after the second pulse of $|1\rangle\xrightarrow{\pi}|3\rangle$
are 
\begin{align}
|\Psi_{\mathrm{II},Q}\rangle=\frac{\mathrm{i}}{\sqrt{2}}(e^{-\mathrm{i}\phi^{Q}_{\mathrm{TL}}}|3\rangle+|2\rangle)
\end{align}
with the corresponding enantioselective $\phi^{Q}_{\mathrm{TL}}$. Specifically, we have $\phi^{L}_{\mathrm{TL}}=\phi$ and $\phi^{R}_{\mathrm{TL}}=\pi+\phi$.
The final states of the two enantiomers after the third pulse of $|2\rangle\xrightarrow{\pi/2}|3\rangle$ are 
\begin{align}
    |\Psi_{\mathrm{III},Q}\rangle=\frac{\mathrm{i}}{2}[(\mathrm{i}+e^{-\mathrm{i}\phi_Q})|3\rangle+(1+\mathrm{i}e^{-\mathrm{i}\phi_Q})|2\rangle].
\end{align}
The overall evolution operator in the basis $\{|1\rangle,|2\rangle,|3\rangle\}$ is 
\begin{equation}
    U_Q=\left(
  \begin{array}{ccc}
    0 & 0 & \mathrm{i}e^{\mathrm{i}\phi^{Q}_{\mathrm{TL}}} \\
    \frac{1}{2}(\mathrm{i}-e^{-\mathrm{i}\phi^Q_{\mathrm{TL}}}) & \frac{1}{2}(1-\mathrm{i}e^{-\mathrm{i}\phi^Q_{\mathrm{TL}}}) &\\
    -\frac{1}{2}(1-\mathrm{i}e^{-\mathrm{i}\phi^Q_{\mathrm{TL}}}) & \frac{1}{2}(\mathrm{i}-e^{-\mathrm{i}\phi^Q_{\mathrm{TL}}}) & 0\\
  \end{array}
\right).
\end{equation}
With the help of $U_Q$, we can give the final states of any initial state. It is known that $U_{L}\ne U_R$ such that the final states of the two 
enantiomers are different, i.e., achieving enantiomeric-specific state transfer. 
The perfect enantiomeric-specific state transfer is achieved when $\phi=\pm\pi/2$. 
When $\phi=\pi/2$, we have 
\begin{align}
    |\Psi_{\mathrm{III},L}\rangle=\mathrm{i}|2\rangle,~~|\Psi_{\mathrm{III},R}\rangle=-|3\rangle.
\end{align}
When $\phi=-\pi/2$, we have $|\Psi_{\mathrm{III},L}\rangle=-|3\rangle$  and $|\Psi_{\mathrm{III},R}\rangle=\mathrm{i}|2\rangle$.


\subsection{The working cavity for 1,2-propanediol}\label{12P}
We use a typical chiral molecule 1,2-propanediol as an example to explicitly show the physical implementation of our working model. 
The rotational constants of 1,2-propanediol are $A=2\pi\times 8.57205$\,GHz, $B=2\pi\times3.6401$\,GHz, and $C=2\pi\times2.79096$\,GHz~\cite{patterson2013enantiomer}. The components of its electric dipole are $|\mu^Q_a|=1.2$\,Debye, $|\mu^Q_b|=1.9$\,Debye, and $|\mu^Q_c|=0.36$\,Debye~\cite{patterson2013enantiomer} with $1\,\mathrm{Debye}\simeq 3.33564\times10^{30}\, \mathrm{C}\cdot \mathrm{m}$. Here, ``$\mathrm{C}$" is short for ``$\mathrm{Coulomb}$" and ``$\mathrm{m}$" is short for ``$\mathrm{meter}$". 
The angular frequencies of the three phase-controlled electromagnetic fields $\bm{E}_{21}$, $\bm{E}_{31}$, and $\bm{E}_{32}$ are 
$\omega_{21}=2\pi\times6.43106$\,GHz, $\omega_{31}=2\pi\times12.21215$\,GHz, and $\omega_{32}=2\pi\times5.78109$\,GHz. $\bm{E}_{21}$, $\bm{E}_{31}$, and $\bm{E}_{32}$ are $Z$-polarized, $Y$-polarized, and $X$-polarized, respectively. The overall phase $\phi$ can be modulated in the regime from $0$ to $2\pi$ by fixing the phases of two fields and changing that of the third one. 

\begin{table}[ht]
\centering
\caption{ The designing of the working cavity with the radius of curvature of the mirror surface is $R_m=40$\,$\mathrm{mm}$ by adjusting the distance between the two mirror centers $d$ and the index of the work mode $q$ to ensure $f_q=5.78109$\,GHz. We also give the corresponding waist of the mode $w_0$, single photon-molecule coupling strength $g_0$, $Q$-factor, and decay rate of cavity $\kappa$.}
\begin{tabular}{cccccccc}
\hline
mode index  & $d~(\mathrm{mm})$  & $w_0~(\mathrm{mm})$ &$g_0~(2\pi\times \mathrm{Hz})$ & $Q$-$\mathrm{factor}~(10^9)$ & $\kappa~(2\pi\times\mathrm{Hz})$&$\Delta f ~(\mathrm{KHz})$ & $\delta f ~(\mathrm{Hz})$\\
\hline
$q = 0$ & 3.460970  &11.5941 &  3.67942 & 0.298502 & 121.686 &$\pm205.68$&   822.7\\
$q = 1$ & 38.638907 & 18.1706 & 0.702688 & 3.33253 & 10.8997&$\pm29.67$  & 118.7\\
$q = 2$ & 73.810999 & 13.7462 & 0.676651 & 6.27981& 5.78419 &$\pm12.68$ & 50.7\\
\hline
\end{tabular}
  \label{Tab1}
\end{table}

We build our cavity based on the spherical Fabry-P\'{e}rot cavity used in {S. Haroche}'s lab~\cite{zhou:tel-00737657}. The radius of curvature of the mirror surface is $R_m=40$\,$\mathrm{mm}$~\cite{zhou:tel-00737657}. By adjusting the distance between the two mirror centers $d$ and the index of the work mode $q$, the angular frequency of the working cavity mode $2\pi\times f_q$ is close to $\omega_{32}=2\pi\times5.78109$\,GHz. The possible cavity geometries are shown in Tab.~\ref{Tab1}. Since $0<d<2R_m$, the geometric escape loss is reduced. 
The mirror substrate in {S. Haroche}'s lab~\cite{zhou:tel-00737657} is made from copper with a very smooth surface. Its surface is covered by a $12$\,$\mathrm{\mu m}$ thick Niobium layer, which becomes superconducting below $9$\,K, such that the surface loss of the mode is eventually reduced, which is denoted by the rate $\kappa_\mathrm{loss}$. 
The diffraction loss can be suppressed by increasing the radius of the mirror, yielding a large enough Fresnel number. The decay rate of the cavity photon is solely determined by the transmissivity of the mirror denoting by $\kappa_T$.  In this sense, we assume the decay rate due to the loss of the photon is $\kappa_\mathrm{loss}=0$, i.e., $\kappa=\kappa_T$. In {S. Haroche}'s lab~\cite{zhou:tel-00737657}, the cavity mode of $f=51$\,GHz with $d=27.6$\,mm has the $Q$-factor of $2.1\times10^{10}$. By using $Q\propto f d$, we give the $Q$-factor of our cavity and the corresponding decay rate $\kappa=2\pi f/Q$ in Tab.~\ref{Tab1}. The frequency of the cavity mode can be finely tuned and precisely
controlled by slightly changing the distance between the two mirrors~\cite{zhou:tel-00737657}. For this purpose, in {S. Haroche}'s lab~\cite{zhou:tel-00737657}, four piezoelectric transducers (PZT) tubes are mounted at the corners of the cavity block and a voltage difference is applied on the PZT tubes to modify the distance $d$, and thus the frequency of the cavity mode. 
In {S. Haroche}'s lab~\cite{zhou:tel-00737657}, the cavity frequency can be tuned by about $\Delta f=\Delta d\cdot \partial f/\partial d=±5$\,MHz with a precision of $\delta f=\delta d \cdot\partial f/\partial d=2.4$\,kHz. By assuming the same tunability of $d$, we give $\Delta f$ and $\delta f$ of our cavity in Tab.~\ref{Tab1}.

\section{The driven Tavis-Cummings model}
In the main text, the enantiomeric hypotheses are decided by monitoring the output of a high finesse Fabry-P\'{e}rot cavity when the molecules pass through the cavity. For a spherical Fabry-P\'{e}rot cavity, the frequency of $\mathrm{TEM}_{q00}$ mode is~\cite{zhou:tel-00737657} 
\begin{align}
f_q=\frac{c}{2d}\left[q+\frac{1}{\pi}\arccos(1-\frac{d}{R_m})\right]
\end{align}
with the velocity of light in vacuum $c$, the distance between the two mirror centers $d$ and the radius of curvature of the mirror surface $R_m$.
The electric field of this mode has a cylindrical symmetry with respect to the cavity axis $Z$, and is given by 
\begin{align}
    \varepsilon_0(\bm{r})= \varepsilon_0 f(r,Z)
\end{align}
with the spatial profile of the mode $f(r,Z)$ $(r=\sqrt{X^2+Y^2})$. In the cylindrical coordinates, it is 
\begin{align}
    f(r,Z)=\frac{w_0}{w(Z)}\cos\left[kZ-\arctan(\frac{\lambda Z}{\pi w^2_0}+\frac{r^2 k}{2 R(Z)})\right]e^{-\frac{r^2}{w(Z)^2}},
\end{align}
where $\lambda=c/f_q$ is the wavelength, $k=2\pi/\lambda$ is the wavenumber. The waist of the mode is 
\begin{align}
    w_0=\left(\frac{\lambda}{2\pi}\sqrt{d(2R_m-d)}\right)^{1/2}.
\end{align}
The radius of curvature and mode width vary with $Z$ as 
\begin{align}
    R(Z)=Z\left[1+\left(\frac{\pi w^2_0}{\lambda Z}\right)^2\right], ~~w(Z)=w_0\sqrt{1+\left(\frac{\lambda Z}{\pi w^2_0}\right)^2}.
\end{align}
The volume of the mode and the amplitude of the electric field associated with one photon are 
\begin{align}\label{MV}
    \mathcal{V}=\int |f(\bm{r})|^2 d^3\bm{r}=\frac{\pi w^2_0 d}{4},~~\varepsilon_0=\sqrt{\frac{\hbar\omega_c}{2\epsilon_0 \mathcal{V}}}
\end{align}
with the dielectric constant in vacuum $\epsilon_0$. 

The angular frequency of $\mathrm{TEM}_{q00}$ mode of the cavity $\omega_c=2\pi f_q$ is close to that of the rotational transition $|2\rangle\leftrightarrow|3\rangle$. 
A classical field at the angular frequency of $\omega_0=\omega_c$ is applied to drive the cavity. In a reference frame rotating with
the driving field, the system is described with the driven Tavis-Cummings model as~\cite{lehmberg1970radiation,plankensteiner2017cavity,plankensteiner2019enhanced,chen2021enantio}  
\begin{align}
    \hat{\mathcal{H}}=\hbar\left[\sum^{N_{m}}_{l=1}\Delta_{m}\hat{\sigma}^{\dagger}_{l}\hat{\sigma}_{l}+\sum^{N_{m}}_{l=1} {g}_l(\hat{c}\hat{\sigma}^{\dagger}_{l}+
    \hat{c}^{\dagger}\hat{\sigma}_{l})+\mathrm{i}\eta(\hat{c}^{\dagger}-\hat{c})\right].
\end{align}
The working mode of the cavity is $\hat{c}$. The molecular operators are defined as $\hat{\sigma}_{l}=|2\rangle_{ll}\langle 3|$ and $\hat{\sigma}^{z}_{l}=2\hat{\sigma}^{\dagger}_{l}\hat{\sigma}_{l}-1$. The cavity-probe and molecule-probe angular frequency detunings are $\Delta_c=\omega_c-\omega_0=0$ and $\Delta_m=\omega_{32}-\omega_0$. The cavity pumping rate is $\eta$. The coupling strength is $ {g}_{l}=g_0f(\bm{r}_l)/2$ with 
\begin{align}
    g_0=\varepsilon_0 \left|\mu^{Q}_b\right|=\sqrt{\frac{2\hbar\omega_c}{\epsilon_0 \pi  d}} \frac{\mu_b}{w_0}.
\end{align} 
The sample propagates in $Y$ direction with forward velocity $\mathrm{v}=1$\,$\mathrm{m/s}$, centering at $X=0$ and $Z=0$ with a cross-sectional area assumed about $L^2\simeq1$\,$ \mathrm{mm^2}$ ($L\simeq 0.1w_0$). We assume that the size of the sample is much smaller than the width of $f(\bm{r})$ (i.e., $L\ll w_0$). 
For simplicity, we have taken the position average of $g$ by 
\begin{align}
   \bar{g}(t)= \bar{g}(\bar{Y})=\frac{g_0}{2S_c} \int^{\prime} f(\bm{r}) dX dZ\simeq  \frac{g_0}{2} \exp\left({-\frac{\bar{Y}^2}{w^2_0}}\right),~~(\bar{Y}=\bar{Y}_0+\mathrm{v}t),
\end{align}
where the subscript ``$\prime$" means the integration is done in the cross-section area $S_c$ of the cavity and the molecular beam. $\bar{Y}_0$ is the averaged initial $Y$ position of the sample. We replace the position-dependent $g_l$ with the averaged $\bar{g}(t)$, yielding 
\begin{align}
    \hat{\mathcal{H}}=\hbar\left[\sum^{N_{m}}_{l=1}\Delta_{m}\hat{\sigma}^{\dagger}_{l}\hat{\sigma}_{l}+\sum^{N_{m}}_{l=1}\bar{g}(\hat{c}\hat{\sigma}^{\dagger}_{l}+
    \hat{c}^{\dagger}\hat{\sigma}_{l})+\mathrm{i}\eta(\hat{c}^{\dagger}-\hat{c})\right].
\end{align} 

\subsection{Mean-field theory in cumulant expansion}
Here, we give more details about our mean-field theory. For this purpose, we first write the equations of motion for operators  
\begin{align}
&\frac{d}{dt}\hat{c}=-\kappa\hat{c}-i\bar{g}\sum^{N_m}_{p=1}\hat{\sigma}_{p}+\eta\nonumber\\
&\frac{d}{dt}\hat{\sigma}_l=-i\Delta_m\hat{\sigma}_l+i\bar{g}\hat{\sigma}^{z}_{l}\hat{c}\nonumber\\
&\frac{d}{dt}\hat{\sigma}^{z}_l=2i\bar{g}(\hat{c}^{\dagger}\hat{\sigma}_{l}-\hat{\sigma}^{\dagger}_{l}\hat{c}).
\end{align}
When averaging the operators, we are confronted with a universal problem in the hybrid system of cavity modes and molecules: averages of operator products (e.g., $\langle\hat{\sigma}^{\dagger}_{l}\hat{c}\rangle$) occur and in general $\langle\hat{\sigma}^{\dagger}_{l}\hat{c}\rangle\ne\langle\hat{\sigma}^{\dagger}_{l}\rangle\langle\hat{c}\rangle$, such that the set
of $c$-number equations is incomplete. This problem will not be solved by deriving equations of motion for those averages, because they will couple to averages of products with more operators. In order to find a full solution, infinitely many equations are needed. 

To solve this problem, we apply some kind of approximation, which allows a cutoff at higher orders and thus makes the set
of $c$-number equations complete. The systematic approach to do this is the so-called cumulant expansion~\cite{kubo1962generalized}. In what follows, we show how to apply the first- and second-order cumulant expansion in the mean-field theory. 

For any operator, it can be divided into its mean value and corresponding fluctuation operator 
\begin{align}
\hat{\mathcal{O}}=\langle\hat{\mathcal{O}}\rangle+d\hat{\mathcal{O}}.
\end{align}
The product of two operators is $\hat{\mathcal{O}}_1 \hat{\mathcal{O}}_2=\langle\hat{\mathcal{O}}_1\rangle\langle\hat{\mathcal{O}}_2\rangle+\langle\hat{\mathcal{O}}_1\rangle d\hat{\mathcal{O}}_2+\langle\hat{\mathcal{O}}_2\rangle d\hat{\mathcal{O}}_1+d\hat{\mathcal{O}}_1 d\hat{\mathcal{O}}_2$.
Because the mean values of fluctuation operator are $zero$ (i.e., $\langle d\mathcal{O}\rangle=0$), we have 
\begin{align}\label{YY1}
\langle\hat{\mathcal{O}}_1 \hat{\mathcal{O}}_2\rangle=\langle\hat{\mathcal{O}}_1\rangle\langle\hat{\mathcal{O}}_2\rangle+\langle d\hat{\mathcal{O}}_1 d\hat{\mathcal{O}}_2\rangle.
\end{align}
For the first order of cumulant expansion, we neglect $\langle d\hat{\mathcal{O}}_1 d\hat{\mathcal{O}}_2\rangle$, yielding 
\begin{align}
&\frac{d}{dt}\langle\hat{c}\rangle=-\kappa\langle\hat{c}\rangle-i\bar{g}\sum^{N_m}_{p=1}\langle\hat{\sigma}_{p}\rangle+\eta,\nonumber\\
&\frac{d}{dt}\langle\hat{\sigma}_l\rangle=-i\Delta_m\langle\hat{\sigma}_l\rangle+i\bar{g}\langle\hat{\sigma}^{z}_{l}\rangle\langle\hat{c}\rangle,\nonumber\\
&\frac{d}{dt}\langle\hat{\sigma}^{z}_l\rangle=2i\bar{g}(\langle\hat{c}^{\dagger}\rangle\langle\hat{\sigma}_{l}\rangle-\langle\hat{\sigma}^{\dagger}_{l}\rangle\langle\hat{c}\rangle).
\end{align}
We note that in our issue the first order of cumulant expansion is enough to obtain the feasible results of our interest. 
To confirm this, we would like to examine the convergence by using the second order of cumulant expansion.

For the second order of cumulant expansion, we use Eq.\,(\ref{YY1}) in the product of three operators $\langle \hat{\mathcal{O}}_1 \hat{\mathcal{O}}_2  \hat{\mathcal{O}}_3\rangle$, yielding
\begin{align}
\langle\hat{\mathcal{O}}_1 \hat{\mathcal{O}}_2  \hat{\mathcal{O}}_3\rangle&=
\langle\hat{\mathcal{O}}_1\rangle\langle\hat{\mathcal{O}}_2\hat{\mathcal{O}}_3\rangle+
\langle\hat{\mathcal{O}}_1\hat{\mathcal{O}}_3\rangle\langle\hat{\mathcal{O}}_2\rangle+
\langle\hat{\mathcal{O}}_1\hat{\mathcal{O}}_2\rangle\langle\hat{\mathcal{O}}_3\rangle
-2\langle\hat{\mathcal{O}}_1\rangle\langle\hat{\mathcal{O}}_2\rangle\langle\hat{\mathcal{O}}_3\rangle+
\langle d\hat{\mathcal{O}}_1d\hat{\mathcal{O}}_2d\hat{\mathcal{O}}_3\rangle.
\end{align}
Similarly, by neglecting $\langle d\hat{\mathcal{O}}_1d\hat{\mathcal{O}}_2d\hat{\mathcal{O}}_3\rangle$, we arrive at the mean-field equations in the second order of cumulant expansion
\begin{align}
&\frac{d}{dt}\langle\hat{c}\rangle=-\kappa\langle\hat{c}\rangle-i\bar{g}N_m\langle\hat{\sigma}_{p}\rangle+\eta,\nonumber\\
&\frac{d}{dt}\langle\hat{\sigma}_l\rangle=-i\Delta_m\langle\hat{\sigma}_l\rangle+i\bar{g}\langle\hat{\sigma}^{z}_{l}\hat{c}\rangle,\nonumber\\
&\frac{d}{dt}\langle\hat{\sigma}^{z}_l\rangle=2i\bar{g}(\langle\hat{c}^{\dagger}\hat{\sigma}_{l}\rangle-\langle\hat{\sigma}^{\dagger}_{l}\hat{c}\rangle),\nonumber\\
&\frac{d}{dt}\langle\hat{\sigma}^{z}_{l}\hat{c}\rangle
=-\kappa\langle\hat{\sigma}^{z}_{l}\hat{c}\rangle+i\bar{g}\langle\hat{\sigma}_{l}\rangle
-i\bar{g}(N_m-1)\langle\hat{\sigma}^{z}_{l}\hat{\sigma}_{p}\rangle+\eta\langle\hat{\sigma}^{z}_{l}\rangle\nonumber\\
&~~~~~~~~~~~~+2i\bar{g}(\langle\hat{c}^{\dagger}\hat{\sigma}_{l}\rangle\langle\hat{c}\rangle+\langle\hat{c}^{\dagger}\rangle\langle\hat{\sigma}_{l}\hat{c}\rangle+\langle\hat{c}^{\dagger}\hat{c}\rangle\langle\hat{\sigma}_{l}\rangle-2\langle\hat{c}^{\dagger}\rangle\langle\hat{\sigma}_{l}\rangle\langle\hat{c}\rangle-\langle\hat{\sigma}^{\dagger}_{l}\rangle\langle\hat{c}\hat{c}\rangle-2\langle\hat{\sigma}^{\dagger}_{l}\hat{c}\rangle\langle\hat{c}\rangle+2\langle\hat{\sigma}^{\dagger}_{l}\rangle\langle\hat{c}\rangle^2),\nonumber\\
&\frac{d}{dt}\langle\hat{\sigma}^{\dagger}_{l}\hat{c}\rangle=
-(\kappa-i\Delta_m)\langle\hat{\sigma}^{\dagger}_l\hat{c}\rangle-i\bar{g}\left[\frac{1}{2}(\langle\hat{\sigma}^{z}_{l}\rangle+1)+(N_m-1)\langle\hat{\sigma}^{\dagger}_{l}\hat{\sigma}_{p}\rangle\right]+\eta\langle\hat{\sigma}^{\dagger}_{l}\rangle
-i\bar{g}(\langle\hat{c}^{\dagger}\rangle\langle\hat{\sigma}^{z}_{l}\hat{c}\rangle+\langle\hat{c}^{\dagger}\hat{\sigma}^{z}_{l}\rangle\langle\hat{c}\rangle+\langle\hat{\sigma}^{z}_{l}\rangle\langle\hat{c}^{\dagger}\hat{c}\rangle\nonumber\\
&~~~~~~~~~~~~-2\langle\hat{c}^{\dagger}\rangle\langle\hat{\sigma}^{z}_{l}\rangle\langle\hat{c}\rangle),\nonumber\\
&\frac{d}{dt}\langle\hat{\sigma}^{\dagger}_{l}\hat{\sigma}_{p}\rangle=-i\bar{g}(\langle\hat{\sigma}^{z}_{l}\rangle\langle\hat{c}^{\dagger}\hat{\sigma}_{p}\rangle+\langle\hat{\sigma}^{z}_{l}\hat{c}^{\dagger}\rangle\langle\hat{\sigma}_{p}\rangle+\langle\hat{\sigma}^{z}_{l}\hat{\sigma}_{p}\rangle\langle\hat{c}^{\dagger}\rangle-2\langle\hat{\sigma}^{z}_{l}\rangle\langle\hat{c}^{\dagger}\rangle\langle\hat{\sigma}_{p}\rangle-\langle\hat{\sigma}^{\dagger}_{l}\rangle\langle\hat{\sigma}^{z}_{p}\hat{c}\rangle-\langle\hat{\sigma}^{\dagger}_{l}\hat{\sigma}^{z}_{p}\rangle\langle\hat{c}\rangle-\langle\hat{\sigma}^{\dagger}_{l}\hat{c}\rangle\langle\hat{\sigma}^{z}_{p}\rangle+2\langle\hat{\sigma}^{\dagger}_{l}\rangle\langle\hat{\sigma}^{z}_{p}\rangle\langle\hat{c}\rangle),\nonumber\\
&\frac{d}{dt}\langle\hat{\sigma}^{z}_{l}\hat{\sigma}_{p}\rangle=-i\Delta_m\langle\hat{\sigma}^{z}_l\hat{\sigma}_p\rangle
+i\bar{g}(\langle\hat{\sigma}^{z}_l\rangle\langle\hat{\sigma}^{z}_{p}\hat{c}\rangle+\langle\hat{\sigma}^{z}_l\hat{\sigma}^{z}_{p}\rangle\langle\hat{c}\rangle+\langle\hat{\sigma}^{z}_l\hat{c}\rangle\langle\hat{\sigma}^{z}_{p}\rangle-2\langle\hat{\sigma}^{z}_l\rangle\langle\hat{\sigma}^{z}_{p}\rangle\langle\hat{c}\rangle)\nonumber\\
&~~~~~~~~~~~~+2i\bar{g}(\langle\hat{c}^{\dagger}\rangle\langle\hat{\sigma}_{l}\hat{\sigma}_{p}\rangle+\langle\hat{c}^{\dagger}\hat{\sigma}_{l}\rangle\langle\hat{\sigma}_{p}\rangle+\langle\hat{c}^{\dagger}\hat{\sigma}_{p}\rangle\langle\hat{\sigma}_{l}\rangle-2\langle\hat{c}^{\dagger}\rangle\langle\hat{\sigma}_{l}\rangle\langle\hat{\sigma}_{p}\rangle
-\langle\hat{\sigma}^{\dagger}_{l}\rangle\langle\hat{c}\hat{\sigma}_{p}\rangle-\langle\hat{\sigma}^{\dagger}_{l}\hat{c}\rangle\langle\hat{\sigma}_{p}\rangle-\langle\hat{\sigma}^{\dagger}_{l}\hat{\sigma}_{p}\rangle\langle\hat{c}\rangle+2\langle\hat{\sigma}^{\dagger}_{l}\rangle\langle\hat{c}\rangle\langle\hat{\sigma}_{p}\rangle),\nonumber\\
&\frac{d}{dt}\langle\hat{\sigma}^{z}_l\hat{\sigma}^{z}_{p}\rangle=4i\bar{g}(\langle\hat{c}^{\dagger}\rangle\langle\hat{\sigma}_{l}\hat{\sigma}^{z}_{p}\rangle+\langle\hat{c}^{\dagger}\hat{\sigma}_{l}\rangle\langle\hat{\sigma}^{z}_{p}\rangle+\langle\hat{c}^{\dagger}\hat{\sigma}^{z}_{p}\rangle\langle\hat{\sigma}_{l}\rangle-2\langle\hat{c}^{\dagger}\rangle\langle\hat{\sigma}_{l}\rangle\langle\hat{\sigma}^{z}_{p}\rangle
-\langle\hat{\sigma}^{\dagger}_{l}\rangle\langle\hat{c}\hat{\sigma}^{z}_{p}\rangle-\langle\hat{\sigma}^{\dagger}_{l}\hat{c}\rangle\langle\hat{\sigma}^{z}_{p}\rangle-\langle\hat{\sigma}^{\dagger}_{l}\hat{\sigma}^{z}_{p}\rangle\langle\hat{c}\rangle-2\langle\hat{\sigma}^{\dagger}_{l}\rangle\langle\hat{c}\rangle\langle\hat{\sigma}^{z}_{p}\rangle),\nonumber\\
&\frac{d}{dt}\langle\hat{c}\hat{c}\rangle=-2\kappa\langle\hat{c}\hat{c}\rangle-i2\bar{g}N_m\langle\hat{\sigma}_{p}\hat{c}\rangle+2\eta\langle\hat{c}\rangle,\nonumber\\
&\frac{d}{dt}\langle\hat{\sigma}_l\hat{c}\rangle=-i\Delta_m\langle\hat{\sigma}_l\hat{c}\rangle+i\bar{g}(\langle\hat{\sigma}^{z}_{l}\rangle\langle\hat{c}\hat{c}\rangle+2\langle\hat{\sigma}^{z}_{l}\hat{c}\rangle\langle\hat{c}\rangle-2\langle\hat{\sigma}^{z}_{l}\rangle\langle\hat{c}\rangle^2)
-\kappa\langle\hat{\sigma}_l\hat{c}\rangle-i\bar{g}(N_m-1)\langle\hat{\sigma}_l\hat{\sigma}_{p}\rangle+\eta\langle\hat{\sigma}_l\rangle,\nonumber\\
&\frac{d}{dt}\langle\hat{c}^{\dagger}\hat{c}\rangle=-2\kappa\langle\hat{c}^{\dagger}\hat{c}\rangle+i\bar{g}N_m(\langle\hat{\sigma}^{\dagger}_{p}\hat{c}\rangle
-\langle\hat{\sigma}_{p}\hat{c}^{\dagger}\rangle)+\eta(\langle\hat{c}\rangle+\langle\hat{c}^{\dagger}\rangle),\nonumber\\
&\frac{d}{dt}\langle\hat{\sigma}_l\hat{\sigma}_p\rangle=-i\Delta_m(\langle\hat{\sigma}_l\hat{\sigma}_p\rangle+\langle\hat{\sigma}_p\hat{\sigma}_l\rangle)
+i2\bar{g}(\langle\hat{\sigma}^{z}_{p}\rangle\langle\hat{c}\hat{\sigma}_l\rangle+\langle\hat{\sigma}^{z}_{p}\hat{\sigma}_l\rangle\langle\hat{c}\rangle+\langle\hat{\sigma}^{z}_{p}\hat{c}\rangle\langle\hat{\sigma}_l\rangle-2\langle\hat{\sigma}^{z}_{p}\rangle\langle\hat{c}\rangle\langle\hat{\sigma}_l\rangle).
\end{align}

\subsection{Negligible of the free space decay and the molecular dipole-dipole interactions}
The equations of motions for mean values of operators $\hat{c}$, $\hat{\sigma}_{l}$, and $\hat{\sigma}^{z}_{l}$ are~\cite{plankensteiner2017cavity,plankensteiner2019enhanced}  
\begin{align}\label{GG}
&\dot{{c}}=-\kappa {c}-\mathrm{i}\sum_{l}\bar{g}{\sigma}_{l}+\eta,\nonumber\\
&\dot{{\sigma}}_{l}=-(\gamma+\mathrm{i}\Delta_m) {\sigma}_{l}+\mathrm{i} \bar{g} {c} \sigma^z_l+\sum_{l^{\prime}\ne l}[\mathrm{i}V(\bm{r}_{ll^{\prime}})+\Gamma(\bm{r}_{ll^{\prime}})] \sigma^z_l {\sigma}_{l^{\prime}},\nonumber\\
&\dot{ {\sigma}}^z_{l}=-2\gamma( \sigma^z_l+1)+2\mathrm{i} \bar{g}( {c}^{ \ast} {\sigma}_{l}- {c} {\sigma}^{ \ast}_{l})-2\sum_{l^{\prime}\ne l}\Gamma(\bm{r}_{ll^{\prime}})( {\sigma}_{l} {\sigma}^{ \ast}_{l^{\prime}}+ {\sigma}^{ \ast}_{l} {\sigma}_{l^{\prime}}),
\end{align}
where we have used the mean values for these operators defined as $c\equiv \langle \hat{c}\rangle $, ${\sigma}_{l}\equiv\langle \hat{\sigma}_{l}\rangle $, and ${\sigma}^z_{l}\equiv\langle \hat{\sigma}^z_{l}\rangle$. $\gamma$ is the free space decay rate. $\Gamma(\bm{r}_{ll^{\prime}})$ and $V(\bm{r}_{ll^{\prime}})$  are the collective decay rate and the strength of collective coherent dipole-dipole interactions due to the vacuum-emitter interaction~\cite{plankensteiner2017cavity,plankensteiner2019enhanced,lehmberg1970radiation}, where we define $\bm{r}_{ll^{\prime}}\equiv\bm{r}_{l}-\bm{r}_{l^{\prime}}$ with the $l$-th molecule's position vector $\bm{r}_{l}$.

The free space decay rate is given as
\begin{align}
\Gamma_0=\frac{1}{4\pi\varepsilon_0}\frac{4\omega^3\mu^2}{3\hbar c^3}.
\end{align}
With the realistic parameters of our scheme given in Sec.\,\ref{12P}, we have the free space decay rates for the three transitions  
$\Gamma_0(2\rightarrow1)=2\pi\times1.8\times10^{-10}$\,Hz, $\Gamma_0(3\rightarrow1)=2\pi\times 3.64\times10^{-11}$\,Hz, and  $\Gamma_0(3\rightarrow2)=2\pi\times 8.06\times10^{-11}$\,Hz. The free space decay rates are sufficiently small. Therefore, the free space decay, the collective decay, and the dipole-dipole interaction
are negligible in our scheme, where the sample's geometry is $L=0.1w_0$ and the particle number of the sample is  $N_m\le 3000$. 

\begin{figure}[ht]
	\centering
	\includegraphics[width=0.8\columnwidth]{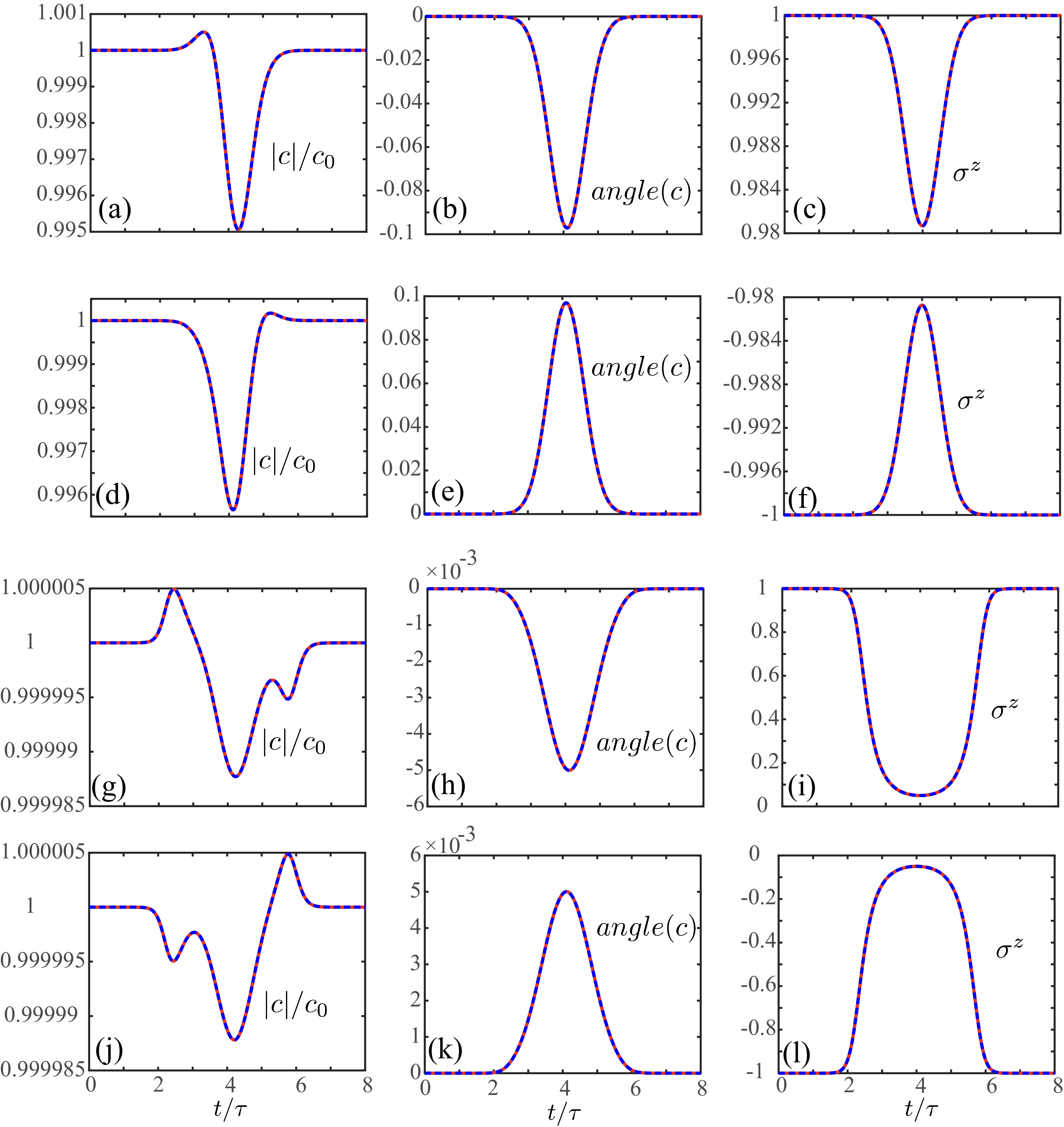}\\
	\caption{{Numerically confirming that the free space decay, the collective decay, and the dipole-dipole interaction are negligible with $N_m=3000$}. Evolutions of the amplitude of $c$ (i.e., $|c|/c(0)$), the phase of $c$ [i.e., $angle(c)$], and $\sigma^z$ based on Eq.~(\ref{SS1}) and Eq.~(\ref{SS3}) in our scheme. (a-c) for the case of $\lambda=0.01$ and $\sigma^z(0)=1$. (d-f) for the case of $\lambda=0.01$ and $\sigma^z(0)=-1$. (g-i) for the case of $\lambda=100$ and $\sigma^z(0)=1$. (j-l) for the case of $\lambda=100$ and $\sigma^z(0)=-1$. The red solid lines are based on Eq.~(\ref{SS1}) without the consideration of the free space decay, the collective decay, and the dipole-dipole interaction. The blue dashed lines are based on Eq.~(\ref{SS3}) with the consideration of the free space decay, the collective decay, and the dipole-dipole interaction.}\label{FigCL}
\end{figure}

Specifically, we consider the working transition of $|2\rangle\leftrightarrow|3\rangle$. The free decay rate is $\gamma=2\pi\times 8.06\times10^{-11}$\,Hz. It is known that~\cite{plankensteiner2017cavity,plankensteiner2019enhanced,lehmberg1970radiation} $\Gamma(\bm{r}_{ll^{\prime}})\le \gamma$ and $\Gamma(\bm{r}_{ll^{\prime}})\to \gamma$ in the case ${r}_{ll^{\prime}}k_m\to 0$ with $k_m=\omega_{32}/c$ with $r_{ll^{\prime}}\equiv|\bm{r}_{ll^{\prime}}| $. For the case with ${r}_{ll^{\prime}}k_m\to 0$, the dipole-dipole interaction strength is~\cite{plankensteiner2017cavity,plankensteiner2019enhanced,lehmberg1970radiation} 
$V(\bm{r}_{ll^{\prime}})\to -{ 3 h\gamma }/{[4(k_m r_{ll^{\prime}})^3]}$ with $h\in[-1,2]$. We adopt the approximation ${r}_{ll^{\prime}}\simeq L N^{-1/3}_m\equiv d$ and thus we have $|V(\bm{r}_{ll^{\prime}})|\le V_{\mathrm{max}}= { 3\gamma N_m}/{[2(k_m L)^3]}$. For the case $N_m=3000$, we have $V_{\mathrm{max}}= 2\pi\times14.4 \times10^{-5}$\,Hz, which is negligible because it is orders of magnitude smaller than $\kappa$, $\eta$, $\Delta_m$, and $g_0$. In this sense, we arrive at the equations of motions 
\begin{align}\label{SS}
&\dot{ {c}}=-\kappa {c}-\mathrm{i}\bar{g}N_m {\sigma}_l+\eta,\nonumber\\
&\dot{ {\sigma}}_l=-\mathrm{i}\Delta_m {\sigma}_l+\mathrm{i}\bar{g}{c} \sigma^z_l,\nonumber\\
&\dot{ {\sigma}}^z_l=2\mathrm{i}\bar{g}( {c}^{ \ast} {\sigma}_l- {c} {\sigma}^{ \ast}_l).
\end{align}
By defining $\sum_l{ {\sigma}}_{l}/N_m\equiv \sigma$ and $\sum_l{ {\sigma}}^z_{l}/N_m\equiv \sigma^z$, we have 
\begin{align}\label{SS1}
&\dot{ {c}}=-\kappa {c}-\mathrm{i}\bar{g}N_m {\sigma}+\eta,\nonumber\\
&\dot{ {\sigma}}=-\mathrm{i}\Delta_m {\sigma}+\mathrm{i}\bar{g}{c} \sigma^z,\nonumber\\
&\dot{ {\sigma}}^z=2\mathrm{i}\bar{g}( {c}^{ \ast} {\sigma}- {c} {\sigma}^{ \ast}).
\end{align}

We would like to further confirm that the free space decay, the collective decay, and the dipole-dipole interaction
are negligible in our scheme by using numerical simulation. Fur this purpose, we replace ${V}(\bm{r}_{ll^{\prime}})$ and $\Gamma(\bm{r}_{ll^{\prime}})$ with $V_{\mathrm{max}}= { 3\gamma N_m}/{[2(k_m L)^3]}$ and $\gamma$, yielding
\begin{align}\label{SS2}
&\dot{ {c}}=-(\kappa+\mathrm{i}\Delta_c) {c}-\mathrm{i}\bar{g}\sum_{l} {\sigma}_{l}+\eta,\nonumber\\
&\dot{ {\sigma}}_{l}=-(\gamma+\mathrm{i}\Delta_m) {\sigma}_{l}+\mathrm{i}\bar{g}{c} \sigma^z_l+\sum_{l^{\prime}\ne l}[\mathrm{i}V_{\mathrm{max}}+\gamma] \sigma^z_l {\sigma}_{l^{\prime}},\nonumber\\
&\dot{ {\sigma}}^z_{l}=-2\gamma( \sigma^z_l+1)+2\mathrm{i}\bar{g}( {c}^{ \ast} {\sigma}_{l}- {c} {\sigma}^{ \ast}_{l})-2\sum_{l^{\prime}\ne l}\gamma( {\sigma}_{l} {\sigma}^{ \ast}_{l^{\prime}}+ {\sigma}^{ \ast}_{l} {\sigma}_{l^{\prime}}).
\end{align}
Then, we have the equation of motions for $\sigma$ and $\sigma^z$ with the consideration of the free space decay, the dipole-dipole interaction, and the collective decay
\begin{align}\label{SS3}
&\dot{ {c}}=-(\kappa+\mathrm{i}\Delta_c) {c}-\mathrm{i}\bar{g}N_m {\sigma}+\eta,\nonumber\\
&\dot{ {\sigma}}=-(\gamma+\mathrm{i}\Delta_m) {\sigma}+\mathrm{i}\bar{g} {c} \sigma^z+(N_m-1)[\mathrm{i}V_{\mathrm{max}}+\gamma] \sigma^z {\sigma},\nonumber\\
&\dot{ {\sigma}}^z=-2\gamma( \sigma^z+1)+2\mathrm{i}\bar{g}( {c}^{ \ast} {\sigma}- {c} {\sigma}^{ \ast})-4(N_m-1)\gamma {\sigma}^{ \ast}{\sigma}.
\end{align}

For simulation, we take $\Delta_c=0$, $\Delta_m=2\pi\times882.7$\,Hz, $\kappa=2\pi\times 121.7$\,Hz, and $g_0=2\pi\times 3.6$\,Hz. The steady state of the cavity in the absence of the molecules is $c=\eta/\kappa$, which means the initial photon number is $N_0=\eta^2/\kappa^2$. The critical photon number for dispersive detection is $N_{\mathrm{cr}} = 4\Delta^2_m/g^2_0\simeq 2.4\times10^5$. Then, we introduce  $N_0=\lambda N_{\mathrm{cr}}$ and the pumping rate is given as  $\eta=\kappa \sqrt{\lambda N_{\mathrm{cr}}}$. In Fig.\,\ref{FigCL}, we give the evolutions of the system based on Eq.~(\ref{SS1}) and Eq.~(\ref{SS3}). The initial values are $c(0)=\sqrt{\lambda N_{\mathrm{cr}}}$ with $\lambda=0.01$ [see Fig.\,\ref{FigCL}(a-f)] and $\lambda=100$ [see Fig.\,\ref{FigCL}(g-l)], $\sigma(0)=0$, and $\sigma^z(0)=\pm1$. Other parameters are $N_m=3000$, $V_{\mathrm{max}}= 2\pi\times14.4 \times10^{-5}$\,Hz, and $\eta=\kappa c(0)$. The lines for the results based on Eq.~(\ref{SS1}) and Eq.~(\ref{SS3}) are coincident. These numerical results further confirm that the free space decay, the dipole-dipole interaction, and the collective decay are negligible in our interested cases. 

Similarly, for transitions $|3\rangle\rightarrow |1\rangle$ and $|2\rangle\rightarrow |1\rangle$, the free space decay, the collective decay, and the dipole-dipole interaction are negligible. We note that the working states of our scheme are in the ground electronic-vibrational state. Because the rotational transition frequencies (of the order of $1\sim 10$\,GHz) and the rotational transition dipoles (of the order of $0.1\sim 1$\,Debye) are typically small for most chiral molecules of interest, such as 1,2-propanediol, menthone, carvone, and 1-indanol. Our conclusion (about the free space decay, the collective decay, and the dipole-dipole interaction) based on the realistic parameters of 1,2-propanediol holds for most chiral molecules of interest in our scheme.


\subsection{Negligible of the collective decay}
\subsubsection{Collective decay in the second-order mean-field theory}
\begin{figure}[ht]
	\centering
	\includegraphics[width=0.8\columnwidth]{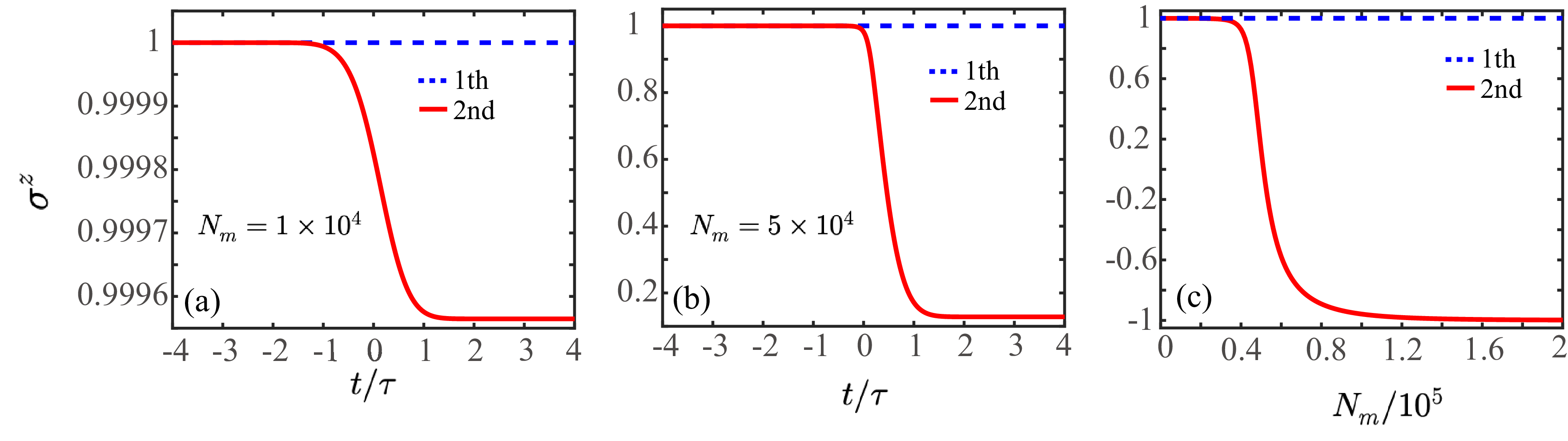}\\
	\caption{Collective decay in the second-order mean-field theory. (a) and (b) give the evolution of $\sigma^z\equiv\sum^{N_m}_{m=1} \langle \hat{\sigma}^z_l\rangle/N_m$ in the initially empty cavity with $\lambda=0$ under the first- (blue dashed lines) and second-order (red solid lines) mean-field theories for $N_m=1\times 10^4$ and $N_m=5\times 10^4$. (c) gives the final $\sigma^z$ at $t=4\tau$ as the function of $N_m$. Other parameters are the same as the main text with $\mathrm{v}=1$\,$\mathrm{m/s}$, $\Delta_c=0$, $\Delta_m=2\pi\times882.7$\,Hz, $\kappa=2\pi\times 121.7$\,Hz, and $g_0=2\pi\times 3.6$\,Hz. The time unit is $\tau=w_0/\mathrm{v}$ with the waist of the cavity mode $w_0$.}\label{FigCD}
\end{figure}

Before going further, we would like to indicate that the key difference between the second- and first-order cumulant
expansion is related to the collective decay of the excited molecules. To illustrate this, we consider the case of the initial empty cavity without the external driving field (i.e., $\lambda=0$). The results are shown in Fig.\,\ref{FigCD}. The results in the second-order mean-field calculations (red solid lines) clearly illustrated that the decay of the excited molecules increases with the molecular numbers, i.e., collective decay. The results in the first-order mean-field calculations (blue dashed lines) show no collective decay.   

\subsubsection{Validity of the first-order mean-field theory in the realistic parameters}
\begin{figure}[ht]
	\centering
	\includegraphics[width=0.8\columnwidth]{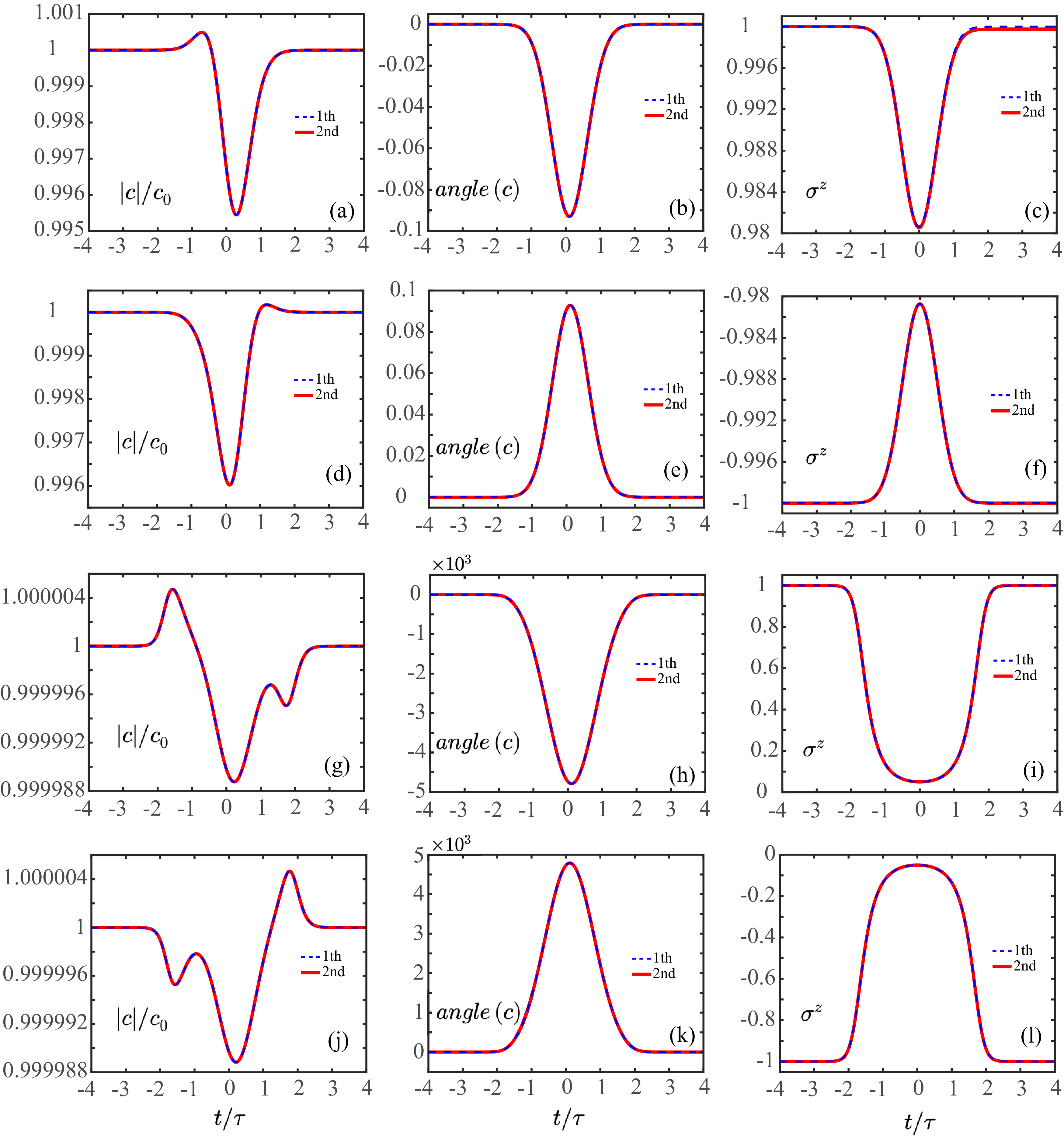}\\
	\caption{Comparing the first- (blue dashed lines) and second-order (red solid lines) results for mean values $\langle \hat{c}\rangle$ and $\langle\hat{\sigma}^z_l\rangle$ (a-f) in the dispersive region with $\lambda=0.01$ and (g-l) beyond the dispersive region with $\lambda=100$. The molecular number is $N_m=3000$. Other parameters are the same as the main text with $\mathrm{v}=1$\,$\mathrm{m/s}$, $\Delta_c=0$, $\Delta_m=2\pi\times882.7$\,Hz, $\kappa=2\pi\times 121.7$\,Hz, and $g_0=2\pi\times 3.6$\,Hz. The time unit is $\tau=w_0/\mathrm{v}$ with the waist of the cavity mode $w_0$.}\label{FigVF}
\end{figure}

In this part, we compare the first- and second-order calculations for mean values $c=\langle \hat{c}\rangle$ and $\sigma=\sum^{N_m}_{m=1}\langle\hat{\sigma}^z_m\rangle/N_m$. The molecular number is chosen as $N_m=3000$. The results in Fig.\,\ref{FigVF} clearly show that under the realistic parameters of our specific system, the results of our interested observables in the first- and second-order mean-field theories are almost the same. This indicates that the collective effects are negligible in our discussions under the realistic parameters of our specific system.

\subsubsection{Applicability of the first-order mean-field theory}
\begin{figure}[ht]
	\centering
	\includegraphics[width=0.8\columnwidth]{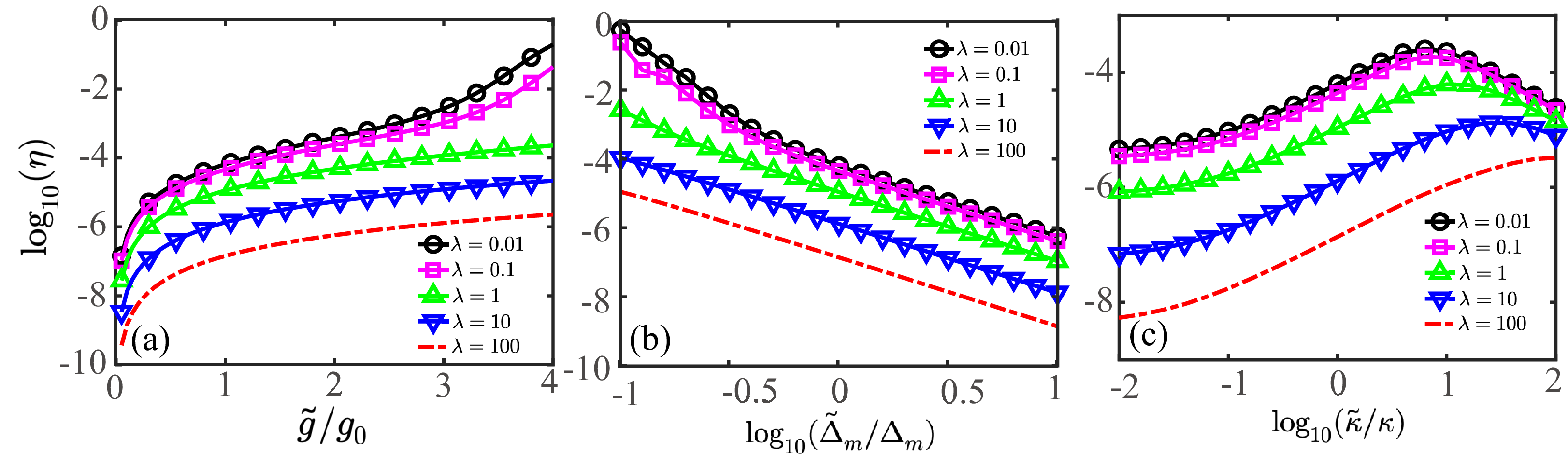}\\
	\caption{Applicability of the first-order mean-field theory. (a-c) give $\eta$ as functions of $\tilde{g}_0$, $\tilde\Delta_m$, and $\tilde\kappa$ for different $\lambda$. In each figure, other parameters are fixed to the values of  $\mathrm{v}=1$\,$\mathrm{m/s}$, $\Delta_c=0$, $\Delta_m=2\pi\times882.7$\,Hz, $\kappa=2\pi\times 121.7$\,Hz, and $g_0=2\pi\times 3.6$\,Hz.}\label{FigPH}
\end{figure}

Further, we would like to give some discussions on the applicability of the first-order mean-field theory. For this purpose, we change the parameters $\tilde{g}_0$, $\tilde\Delta_m$, and $\tilde\kappa$, where ``$\tilde{~~}$'' is used to distinguish them from the values ${g}_0$, $\Delta_m$, and $\kappa$ in the main text. In our discussions, we are interested in the mean values $\bar{n}$, where we have $t_0=-4\tau$ and $t_f=4\tau$. The time unit is $\tau=w_0/\mathrm{v}$ with the waist of the cavity mode $w_0$. The molecules are all initially in the excited state with $\sigma^z=1$. To show the applicability of the first-order mean-field theory, we define 
\begin{align}
\eta=\left|\frac{\bar{n}_{1st}-\bar{n}_{2nd}}{\bar{n}_{1st}+\bar{n}_{2nd}}\right|,
\end{align}
where the subscripts $1st$ and $2nd$ denote the results of first- and second-order mean-field theories. In Fig.\,\ref{FigPH}, we show $\eta$ as functions of $\tilde{g}_0$, $\tilde\Delta_m$, and $\tilde\kappa$ for different $\lambda$. It clearly shows that the first-order mean-field theory is applicable to wide ranges. In fact, most chiral molecules of interest can work in such a region, because we can appropriately choose the working states of them. 

\section{Single-shot decision of molecular chirality: Balanced Homodyne detection}
\subsection{Balanced Homodyne detection}
The homodyne detection is done by applying a strong local field with the same frequency as the cavity's output field. 
By using the input-output relation of the cavity, we have the output field 
\begin{align}
    \sqrt{\frac{\kappa}{2}}{c}={c}_{\mathrm{out}}.
\end{align}
The local field ${c}_{\mathrm{lo}}$ and the output field ${c}_{\mathrm{out}}$ are mixed by $50:50$ beam splitter, yielding two modes 
\begin{align}
    {c}_{\pm}=\frac{1}{\sqrt{2}}({c}_{\mathrm{lo}}\pm {c}_{\mathrm{out}}).
\end{align}  
The two modes are detected and the results are compared to give a measurement equaling to 
\begin{align}
{N}={N}_{+}-{N}_{-}={c}^{\ast}_{\mathrm{lo}}{c}_{\mathrm{out}}+c.c. \simeq\sqrt{2\kappa N_{\mathrm{lo}}}|{c}|\cos(\varphi-\varphi_{\mathrm{lo}}),
\end{align}
where the incident rate of the local field is $N_{\mathrm{lo}}=\left|{c}_{\mathrm{lo}}\right|^2$. The phase of the cavity field is $\varphi$. The phase of the local field is $\varphi_\mathrm{lo}$.
The overall signal collected in a time span of $(0,t)$ is given by 
\begin{align}\label{nbar}
   \bar{n}=\int^{t}_{0}{N}(t^{\prime})dt^{\prime}.
\end{align} 
For the measurements of $\bar{n}_{\pm}\equiv\int^{t}_0 N_{\pm}(t^{\prime})dt^{\prime}\simeq N_{\mathrm{lo}}t/2$, the variances due to the quantum noise are given as~\cite{horak2003possibility} 
\begin{align}\label{delta}
    \delta^2_{\pm}\simeq \frac{1}{2}N_{\mathrm{lo}}t,
\end{align}
yielding the variance of $\bar{n}$~\cite{horak2003possibility}  
\begin{align}
\delta^2=\delta^2_{+}+\delta^2_{-}=N_{\mathrm{lo}}t.
\end{align}
Then, in the case of $N_{\mathrm{lo}}t\gg1$, a single-shot measurement in the duration $(0,t)$ yields a random variable, whose probability density function is approximate to Gaussian distribution of 
\begin{align}\label{FN}
    F(n)\equiv\frac{1}{\sqrt{2\pi\delta^2}}\exp\left[\frac{-(n-\bar{n})^2}{2\delta^2}\right].
\end{align}
In Ref.\,\cite{horak2003possibility}, the probability density functions of $\bar{n}_{\pm}$ are assumed to be Poisson distribution, which is approximate to Gaussian distribution in the case of $N_{\mathrm{lo}}t\gg1$. 
We take such an approximation and work out the distribution corresponding to $\bar{n}$ as shown in Eq.\,(\ref{FN}).

\subsection{Physical mechanism of single-shot determination of molecular chirality}
In the dispersive case, we model the evolution of ${c}$ as 
\begin{align}
    {c}_{Q}(t)=|c(0)|\exp[-\mathrm{i}\varphi_{Q}(t)]
\end{align}
with $Q=L,R$ denoting the two enantiomeric hypotheses. 
The overall signals without the appearance of molecules are 
\begin{align}
    &\bar{n}_{N}=\sqrt{2\kappa N_{\mathrm{lo}}}|c(0)|\int^{t}_{0}\cos(\varphi_{\mathrm{lo}})dt^{\prime}.
\end{align}
For the two hypotheses $\varphi_L=-\varphi_R=\varphi$, we use $|\varphi(t)|\ll0$, yielding 
\begin{align}
    &\bar{n}_{L}\simeq\bar{n}_{N}- \sqrt{2\kappa N_{\mathrm{lo}}}|c(0)|\int^{t}_{0}\varphi(t^{\prime})\sin(\varphi_{\mathrm{lo}})dt^{\prime},\nonumber\\
    &\bar{n}_{R}\simeq\bar{n}_{N}+ \sqrt{2\kappa N_{\mathrm{lo}}}|c(0)|\int^{t}_{0}\varphi(t^{\prime})\sin(\varphi_{\mathrm{lo}})dt^{\prime}.
\end{align}
We note that $\bar{n}_L$ and $\bar{n}_R$ are well separated when $\sin(\varphi_{\mathrm{lo}})=\pm1$. In this sense, 
\begin{align}
    \bar{n}_{L}=-\bar{n}_{R}=\bar{n},~~|\bar{n}|=\sqrt{2\kappa N_{\mathrm{lo}}}\left|c(0)\int^{t}_{0}\varphi(t^{\prime})dt^{\prime}\right|.
\end{align}

We label the mean values $\bar{n}$ in the two enantiomeric hypotheses as $\bar{n}_L$ and $\bar{n}_R$.
After a single-shot measurement, the decision is made as follows. We assume that $\bar{n}_L>0$. If $n>n_c$, we decide the molecule is left-handed. If $n<-n_c$, we decide the molecule is right-handed. Otherwise, we discard the result and such a single-shot measurement fails to tell the molecular chirality. The conditional probabilities of outcomes ``yes" and ``no" given in the two hypotheses are
\begin{align}
    &P(\mathrm{yes}|H_L)=\int^{+\infty}_{+n_c}F_{L}(n)dn,~~P(\mathrm{no}|H_L)=1-P(\mathrm{yes}|H_L),\nonumber\\
    &P(\mathrm{yes}|H_R)=\int^{-n_c}_{-\infty}F_{R}(n)dn,~~P(\mathrm{no}|H_R)=1-P(\mathrm{yes}|H_R).
\end{align}
$F_{Q}$ is the probability density function. By assuming the equally-likely hypotheses, we give the error probability as 
\begin{align}
    P_{\mathrm{err}}=\frac{1}{2}\sum_{Q=L,R}P(\mathrm{no}|H_Q)
\end{align}
In the case of $\bar{n}_L=-\bar{n}_R$, the minimum $P_{\mathrm{err}}$ with respect to the choice of $n_c$ is achieved when 
\begin{align}
    n_c=0.
\end{align}  
Then, the single-shot error probability is~\cite{barzanjeh2015microwave} 
\begin{align}
    P_{\mathrm{err}}=\frac{1}{2}\mathrm{Erfc}(\frac{ \mathrm{SNR}}{\sqrt{2}}),
\end{align}
where $\mathrm{Erfc}(x)$ is the complementary error function and the signal-to-noise rate is 
\begin{align}\label{SNR}
    \mathrm{SNR}=\frac{|\bar{n}_L-\bar{n}_R|}{2\delta}=\sqrt{{2\kappa N_0}t} \bar{\varphi}
\end{align}
with the absolute value of the mean dispersive phase shift 
\begin{align}\label{AMP}
    \bar{\varphi}\equiv\frac{1}{t}\left|\int^{t}_{0}\varphi(t^{\prime})dt^{\prime}\right|.
\end{align}

\subsection{Analytical results in the dispersive limit}
From Eq.~(\ref{SNR}), we know that the opening and closing of the detector should be optimized to obtain a good signal-to-noise rate. Here, we would like to address this issue analytically in the dispersive limit, where we have
\begin{align}
\sigma^z(t)\simeq \sigma^z(0)\equiv \sigma^z(0).
\end{align} 
We assume the sample moves sufficiently slowly and thus the hybrid system stays in its instantaneous ``steady state" governed by the following equations
\begin{align}\label{SSE}
&0=-\kappa {c}-\mathrm{i}\bar{g}N_m {\sigma}+\eta,\nonumber\\
&0=-\mathrm{i}\Delta_m {\sigma}+\mathrm{i}\bar{g}{c} \sigma^z(0).
\end{align}
The second equation gives 
\begin{align}
    \sigma=\frac{\bar{g}c\sigma^z}{\Delta_m}.
\end{align} 
Inserting it into the first equation, we arrive at 
\begin{align}
c=\frac{\eta}{\kappa +\mathrm{i}\bar{g}^2N_m {\sigma^z}{\Delta^{-1}_m}}\simeq \frac{\eta}{\kappa}e^{i\varphi},~~\kappa\gg|\bar{g}^2N_m {\sigma^z}/\Delta_m|
\end{align}
where we have 
\begin{align}\label{SB1}
\varphi=-\frac{\sigma^z(0)\bar{g}^2N_m}{\kappa\Delta_m}
\end{align}

\subsection{Optimization of detection for moving samples}\label{ODMS}
The sample moves from $\bar{Y}_0=-4 w_0$ to $\bar{Y}_t=4 w_0$. The signals are collected from $t_0=-\bar{Y}_0/\mathrm{v}-M_Y\tau$ to $t_f=-\bar{Y}_0/\mathrm{v}+M_Y\tau$ with $\bar{Y}_0/\mathrm{v}=-4\tau$. The absolute value of the mean dispersive phase shift is
\begin{align}
    \bar{\varphi}= \left|\frac{1}{\kappa \Delta_m t}\int^{t_f}_{t_0} \sigma^z(\tau)[\bar{g}(\tau)]^2d\tau\right|= \frac{\sqrt{\pi}}{8M_Y\sqrt{2}}\mathrm{Erf}(\sqrt{2}M_{Y})\frac{ g^2_0N_m}{\kappa |\Delta_m| },
\end{align}
Then we have 
\begin{align}
    \mathrm{SNR}=\frac{|\bar{n}_L-\bar{n}_R|}{2\sigma}= \frac{\sqrt{{\kappa N_0} w_{0}\pi}}{4\sqrt{2}\sqrt{\mathrm{v}M_Y}}\mathrm{Erf}(\sqrt{2}M_{Y})\frac{ g^2_0 N_m}{\kappa |\Delta_m| }.
\end{align}
The maximum signal-to-noise rate with respect to $M_Y$ is achieved at $M_Y\simeq 0.7$, such that 
\begin{align}\label{SNR1}
    \mathrm{SNR}\simeq\sqrt{\frac{{ N_0} w_{0}\pi}{2\kappa \mathrm{v}}}\frac{ N_m g^2_0}{ 4\Delta_m }=\frac{g_0}{2}\sqrt{\frac{w_0 \pi}{\kappa}}\sqrt{\frac{\lambda}{2\mathrm{v}}}N_m.
\end{align}
Because $w_0 $, $\kappa$, and $g_0$ are functions of $q$ (see Tab.\ref{Tab1}), we define
\begin{align}
  \mathcal{F}(q)\equiv\frac{g_0}{2}\sqrt{\frac{w_0 \pi}{\kappa}}.
\end{align}
According to Tab.~\ref{Tab1}, we have 
$\mathcal{F}(0)=8.01039\times10^{-2}$, $\mathcal{F}(1)=6.37345\times10^{-2}$, and  $\mathcal{F}(2)=7.32773\times10^{-2}$ in the unit of $\sqrt{\mathrm{m\cdot Hz}}$. Therefore, we would like to choose the cavity geometry with respect to $q=0$ in Tab.~\ref{Tab1}. The other parameters $\lambda$ and $\mathrm{v}$ in Eq.~(\ref{SNR1}) are tunable. Under the optimized geometry, the numerical result of the signal-to-noise ratio corresponding to the numerical simulation in Fig.~\ref{FigCL} is $\mathrm{SNR}=16.2$. The numerical result of the signal-to-noise rate is given by numerically solving Eqs.~(\ref{nbar}),(\ref{delta}), and (\ref{SNR}). The analysis result according to Eq.~(\ref{SNR1}) for the parameters of Fig.~\ref{FigCL} is $\mathrm{SNR}=16.8$, which shows a good agreement with the numerical result of $\mathrm{SNR}=16.2$. 



\section{Chiral discrimination of trapped single molecule}

\subsection{Analytical results in the dispersive limit}\label{TAR}
For the trapped case, we have $\bar{g}_{\mathrm{trap}}=g_0/2$ is time-independent. In the dispersive case, we have 
\begin{align}
c_{\mathrm{trap}}=\frac{\eta}{\kappa +\mathrm{i}\bar{g}_{\mathrm{trap}}^2N_m {\sigma^z}{\Delta^{-1}_m}}\simeq \frac{\eta}{\kappa}e^{i\varphi_{\mathrm{trap}}},
\end{align}
where we have 
\begin{align}\label{SB2}
\varphi_{\mathrm{trap}}=-\frac{\sigma^z{g}^2_{0}N_m}{4\kappa\Delta_m}.
\end{align}

By using Eqs.~(\ref{SNR}) and (\ref{AMP}), we have 
\begin{align}\label{SNR11}
    \mathrm{SNR}
    = {g}_{0}N_m\sqrt{\frac{{ \lambda }t}{2{\kappa  }}}.
\end{align}
To obtain the single-molecule chiral discrimination, the molecule should be trapped in the cavity for a sufficiently long time with 
\begin{align}
t\ge t_c=\frac{18\kappa}{{g}^{2}_{0}\lambda}\equiv \lambda^{-1}t_{0},
\end{align}
where $t_c$ corresponds to the case with $\mathrm{SNR}=3$ and $t_0\equiv{18\kappa}/{{g}^{2}_{0}}\simeq 26$\,s for the case of the $q=0$ cavity in Tab.\,\ref{Tab1}. Thus, for the typical dispersive region with $\lambda=0.01$, the molecule should be trapped in the cavity with $t\ge 2600$\,s for determining its chirality.


\subsection{Numerical simulations in and beyond the dispersive regions}
\begin{figure}[ht]
	\centering
	\includegraphics[width=0.95\columnwidth]{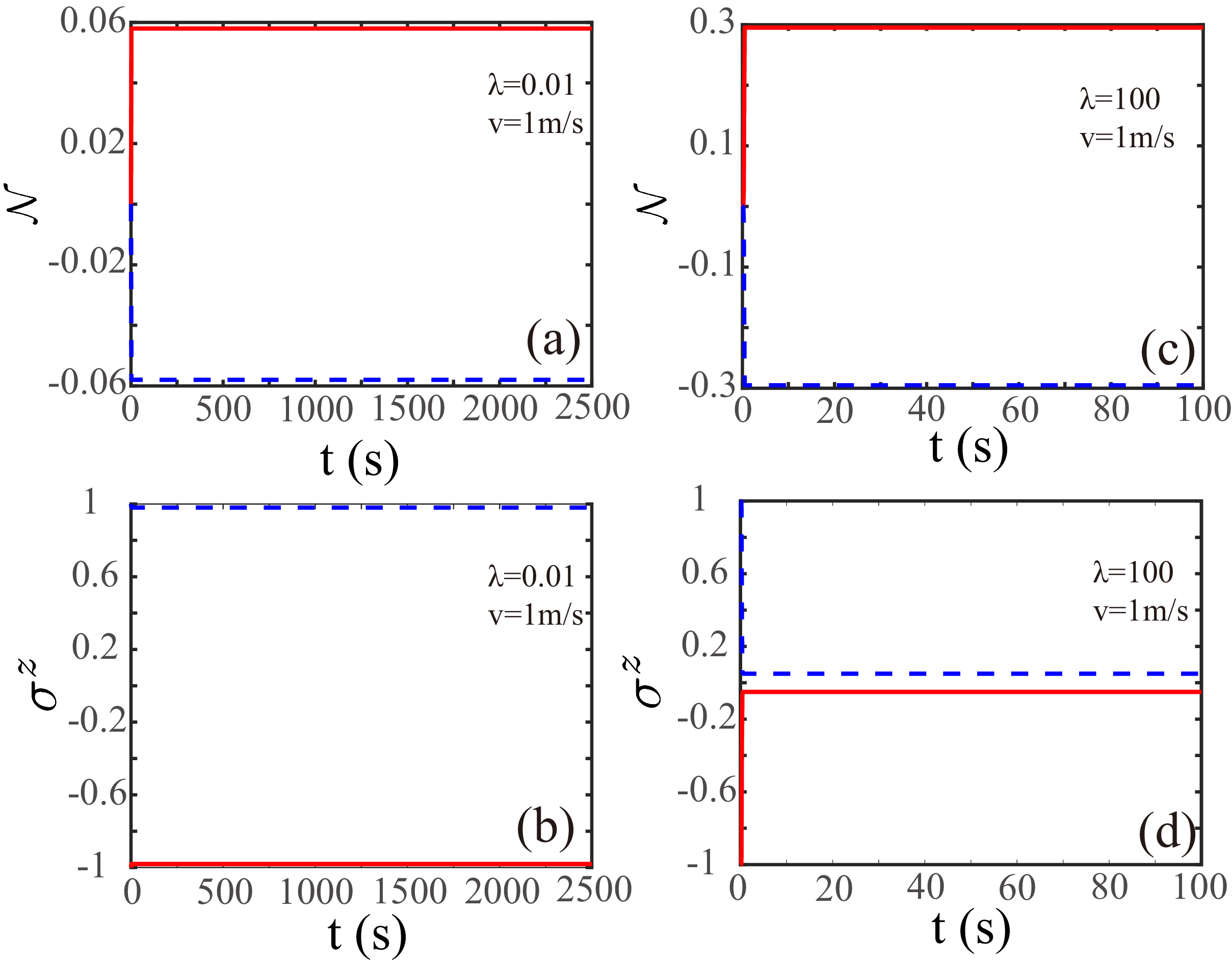}\\
	\caption{Enantioselective responses of the hybrid cavity-molecule system for the case, where the molecule enters the cavity with forward velocity $\mathrm{v}$ and then is trapped in the cavity for a sufficiently long time. (a-c) in the dispersive region with $\lambda=0.01$ and (d-f) beyond the dispersive region with $\lambda=100$. The left- and right-handed molecules are marked with blue and red colors.}\label{FigS3}
\end{figure}

In Fig.\,\ref{FigS3}, we give the numerical simulations for the case, where the molecule enters the cavity with forward velocity $\mathrm{v}=1$\,m/s and then is trapped in the cavity for a sufficiently long time. In Figs.\,\ref{FigS3}(a) and (b) show the results for the dispersive case with $\lambda=0.01$. As we have shown in Sec.\,\ref{TAR}, $t\ge 2600$\,s is needed to make $\mathrm{SNR}\ge3$. Then, the simulation time is about $2000$\,s in Figs.\,\ref{FigS3}(a) and (b). The system quickly enters a steady state. The detection starts at $t_0=0$ and ends at $t_f$. Then, the signal-to-noise rates are 
\begin{align}
\mathrm{SNR}=\sqrt{t_f}\mathcal{N}_s,
\end{align}
where $\mathcal{N}_s$ is the value of $\bar{N}$ at the steady state. For the dispersive case, we find that $\mathcal{N}_s\simeq0.06$\,$\sqrt{\mathrm{Hz}}$. The critical time for $\mathrm{SNR}\ge 3$ is $t_f\simeq 2500$\,s. 
For the non-dispersive case, we find that $\mathcal{N}_s\simeq0.3$\,$\sqrt{\mathrm{Hz}}$, which means the critical time for $\mathrm{SNR}\ge 3$ is $t_f\simeq 100$\,s.  Although the final state in the non-dispersive case is changed, the molecule is still in the electronic-vibrational ground state, and thus the chiral molecules are not destroyed or disturbed after detection. 

\bibliography{ycref}